\documentclass[]{jfm}

\usepackage{graphicx}
\usepackage{epstopdf,epsfig}
\usepackage{newtxtext}
\usepackage{newtxmath}
\usepackage{natbib}
\usepackage{hyperref}
\hypersetup{
    colorlinks = true,
    urlcolor   = blue,
    citecolor  = black,
}

\newcommand{\RomanNumeralCaps}[1]
\nolinenumbers

\usepackage{booktabs}
\usepackage{color,soul}

\usepackage{cancel}

\usepackage{float}

\usepackage{array}
\newcolumntype{L}[1]{>{\raggedright\let\newline\\\arraybackslash\hspace{0pt}}m{#1}}
\newcolumntype{C}[1]{>{\centering\let\newline\\\arraybackslash\hspace{0pt}}m{#1}}
\newcolumntype{R}[1]{>{\raggedleft\let\newline\\\arraybackslash\hspace{0pt}}m{#1}}

\def\ba#1\ea{\begin{align}#1\end{align}}
\def\bsa#1#2\esa{\begin{subequations}\label{#1}
\begin{align}#2\end{align} \end{subequations}}

\def\lp{\left(}\def\rp{\right)}\def\lb{\left[}\def\rb{\right]}

\allowdisplaybreaks

\def\f{\frac}
\def\p{\partial}

\def\bs{\boldsymbol}
\def\NA{\bs{\nabla}}
\def\DEL{\bs{\nabla^2}}
\def\u{{\bs{u}}}
\title[Competition between {Rayleigh--Bénard} and horizontal convection]{Competition between {Rayleigh--Bénard} and horizontal convection}

\author[L.-A. Couston, J. Nandaha, B. Favier]{Louis-Alexandre Couston\aff{1}
  \corresp{\email{louis.couston@ens-lyon.fr}}, 
  Joseph Nandaha\aff{1}, 
  Benjamin Favier\aff{2}}

\affiliation{\aff{1}ENSL, UCBL, CNRS, Laboratoire de physique, F-69342 Lyon, France \aff{2}Aix Marseille Univ, CNRS, Centrale Marseille, IRPHE UMR 7342, Marseille, France}

\begin{document}
\maketitle

\begin{abstract}

We investigate the dynamics of a fluid layer subject to an imposed bottom heat flux and a top monotonically-increasing temperature profile driving horizontal convection. We use direct numerical simulations and consider a large range of flux-based Rayleigh numbers $10^6 \leq Ra_F \leq 10^9$ and imposed top horizontal to bottom vertical heat flux ratios $0 \leq \Lambda \leq 1$. The fluid domain is a closed two-dimensional box with aspect ratio $4\leq \Gamma \leq 16$ and we consider no-slip boundaries and adiabatic side walls. We demonstrate a regime transition from {Rayleigh--Bénard} convection (RB) to horizontal convection (HC) at $\Lambda\approx 10^{-2}$, which is independent of $Ra_F$ and $\Gamma$. At small $\Lambda$, the flow is organized in multiple overturning cells with approximately unit aspect ratio, while at large $\Lambda$ a single cell is obtained. The RB-relevant Nusselt number scaling with $Ra_F$ and the HC-relevant Nusselt number scaling with the horizontal Rayleigh number $Ra_L=Ra_F\Lambda\Gamma^4$ are in good agreement with previous results from classical RB convection and HC studies in the limit $\Lambda \ll 10^{-2}$ and $\Lambda \gg 10^{-2}$, respectively. We demonstrate that the system is multi-stable near the transition $\Lambda\approx10^{-2}$, i.e. the exact number of cells not only depends on $\Lambda$ but also on the system's history. Our results suggest that subglacial lakes, which motivated this study, are likely to be dominated by RB convection, unless the slope of the ice-water interface, which controls the horizontal temperature gradient via the pressure-dependence of the freezing point, is greater than unity.

\end{abstract}



\section{Introduction}\label{sec:intro}

Buoyancy-driven flows are ubiquitous in nature and industrial processes. Geothermal heating generates plumes in Earth's mantle that drive plate tectonics \citep{Jaupart2007}, variations of solar irradiance with latitude drive large-scale winds in Earth's atmosphere \citep{Trenberth2001}, brine rejection from sea ice increases the density of near-surface water masses that contribute to the large-scale ocean circulation \citep{Jacobs2004,Abernathey2016}, and metal assembly or coating through laser-induced heat deposition involve multi-component phase changes and fluid convection that can affect material microstructure \citep{Gan2017}. 

{Rayleigh--Bénard} (RB) convection and horizontal convection (HC) are two canonical buoyancy-driven flow configurations that have attracted significant interest. RB convection considers fluid motions between horizontal plates held at different temperatures, such that the diffusive base state is gravitationally unstable \citep{Ahlers2009}, whereas HC considers an inhomogeneous heat flux or temperature distribution along a single horizontal boundary, which is baroclinically unstable and drive a vigorous boundary-layer flow \citep{Hughes2008}. The behavior of RB convection and HC is typically examined through the scaling of the Reynolds number $Re$, which is a proxy for fluid velocity, and Nusselt number $Nu$, which is a proxy for the efficiency of heat transport by convection, with the control parameters, including notably, the Rayleigh number $Ra$, which measures the available potential energy relative to dissipation processes. Research over the past few decades {has seen the development of} detailed phase diagrams for the flow regimes and scalings of $Re$ and $Nu$ as functions of $Ra$, as well as the Prandtl number $Pr$, which compares viscosity to thermal diffusivity, for both RB convection \citep{Grossmann2000,Ahlers2009} and HC \citep{Mullarney2004,Shishkina2016}. However, many open questions remain. For instance, recent studies on RB convection aim to address the existence of the ultimate regime \citep{Zhu2018} or its large-scale organization \citep{Pandey2018,Wang2020,Vieweg2021}, while active topics of research in HC include the emergence and properties of turbulence, which is spatially heterogeneous, the definition of a thermodynamically-compelling Nusselt number, which is not as straightforward as in RB convection  \citep{Paparella2002,Scotti2011,Gayen2014,Passaggia2017,Rocha2020} and the effect of rotation \citep{Vreugdenhil2019,Gayen2022}. 

The fluid dynamics resulting from the superposition of RB convection with HC has received surprisingly limited attention, in spite of being of fundamental interest and having potentially important applications in the environment. For instance, RB convection and HC may concomitantly control the fluid dynamics within subglacial lakes in Greenland and Antarctica, which impact the dynamics of ice sheets and most likely host extremophiles of interest to astrobiology \citep{Cockell2011,Livingstone2022}. Subglacial lakes{, which are typically fresh except when active or close to grounding lines \citep{Priscu2021},} are exposed to geothermal heating as well as horizontal temperature gradients along the ice-water interface when the ice thickness above is spatially variable. Thicker ice produces larger pressure at the ice-water interface and thus lower interface temperature, because the freezing (or fusion) temperature of water decreases with pressure \citep{Thoma2010a}. The fluid dynamics of planetary oceans may also be affected by both RB convection and HC, as oceans receive solar radiations that vary with latitude and typically experience geothermal heating \citep{Wang2016}. Another field where thermal convection might be affected by boundary inhomogeneities in a direction transverse to gravity is the Earth's liquid outer core. Heterogeneous heat fluxes along the core-mantle boundary, which are due to large-scale convective patterns within the mantle, can sustain large-scale azimuthal flows and affect heat fluxes across the fluid layer \citep{Sumita1999,Mound2017}. 

Past studies of the dual {Rayleigh--Bénard}-Horizontal (RBH) configuration include a handful of simulations and experiments relevant to subglacial lakes and open oceans. RBH dynamics underly a series of realistic numerical simulations of subglacial lakes that used a large-scale ocean code with parameterized subgrid-scale processes \citep{Thoma2007,Thoma2009a} and a laboratory experiment of lake Vostok \citep{Wells2008}. There is yet no consensus on the type of fluid motions expected in subglacial lakes: \citet{Thoma2007,Thoma2009a} predict HC-driven large-scale circulations in lakes Vostok and Concordia affected by rotation; \citet{Wells2008} found that the dynamics in lake Vostok is better described by rotating RB convection, i.e. dominated by multiple columnar vortices; and, \citet{Couston2021b} predicts non-rotating RB convection in most subglacial lakes. In open ocean research, several studies \citep{Hofmann2009} have shown using General Circulation Models (GCMs) that the geothermal flux affects the global ocean circulation, which is otherwise primarily driven by winds and heat fluxes at the air-sea interface. The impact of geothermal heating on the Meridional Overturning Circulation (MOC) of the Atlantic Ocean has also been investigated through idealized numerical simulations \citep{Mullarney2006} and laboratory experiments \citep{Wang2016}, wherein the MOC is driven by a horizontally-varying source of buoyancy or buoyancy flux. Both studies focused on dynamical regimes dominated by HC but demonstrated significant effects of bottom heating: in a box with aspect ratio $\Gamma=6$, \citet{Mullarney2006} found that the volume flux driven by HC is 145\% higher with a bottom heat flux equal to just 10\% the amount of heat extracted from the top boundary, while in a box with aspect ratio $\Gamma=1$, \citet{Wang2016} found a 260\% increase of the volume flux with bottom heating equal to 6.8\% the amount of heat input through half of the top boundary.

In this paper, our goal is to identify and characterize the transition from RB convection to HC as a function of the control parameters, including, most importantly, the ratio of the imposed horizontal temperature gradient $\lambda$ along the top boundary multiplied by  the thermal conductivity $k$ (which yields a horizontal heat flux), to the bottom heat flux $F$, which we write as $\Lambda=k\lambda/F$. To this end, we run a large number of two-dimensional numerical simulations with variable bottom and surface forcing (listed in table \ref{tab:sims}), and diagnose the resulting dynamics through the Reynolds and Nusselt numbers, as well as the characteristic length scale of overturning motions (see \S~\ref{sec:phen}-\ref{sec:auto}). We also run a set of simulations over very long time scales (tens of diffusive time scales) near the transition, in order to demonstrate that RBH convection is multi stable for some parameters (\S~\ref{sec:hyst}). The manuscript is organized as follows. In section \S~\ref{sec:pbm} we introduce the dimensional and dimensionless governing equations as well as the numerical method. In section \S~\ref{sec:res} we show results highlighting the regime transition between RB-like convection at $\Lambda\ll 10^{-2}$ and horizontal convection at $\Lambda\gg 10^{-2}$, and we demonstrate the existence of multiple flow states for the same set of problem parameters. Finally, we conclude in \S~\ref{sec:conc}.

\section{Problem formulation}\label{sec:pbm}

\begin{figure}
\centering
\includegraphics[width=0.45\textwidth]{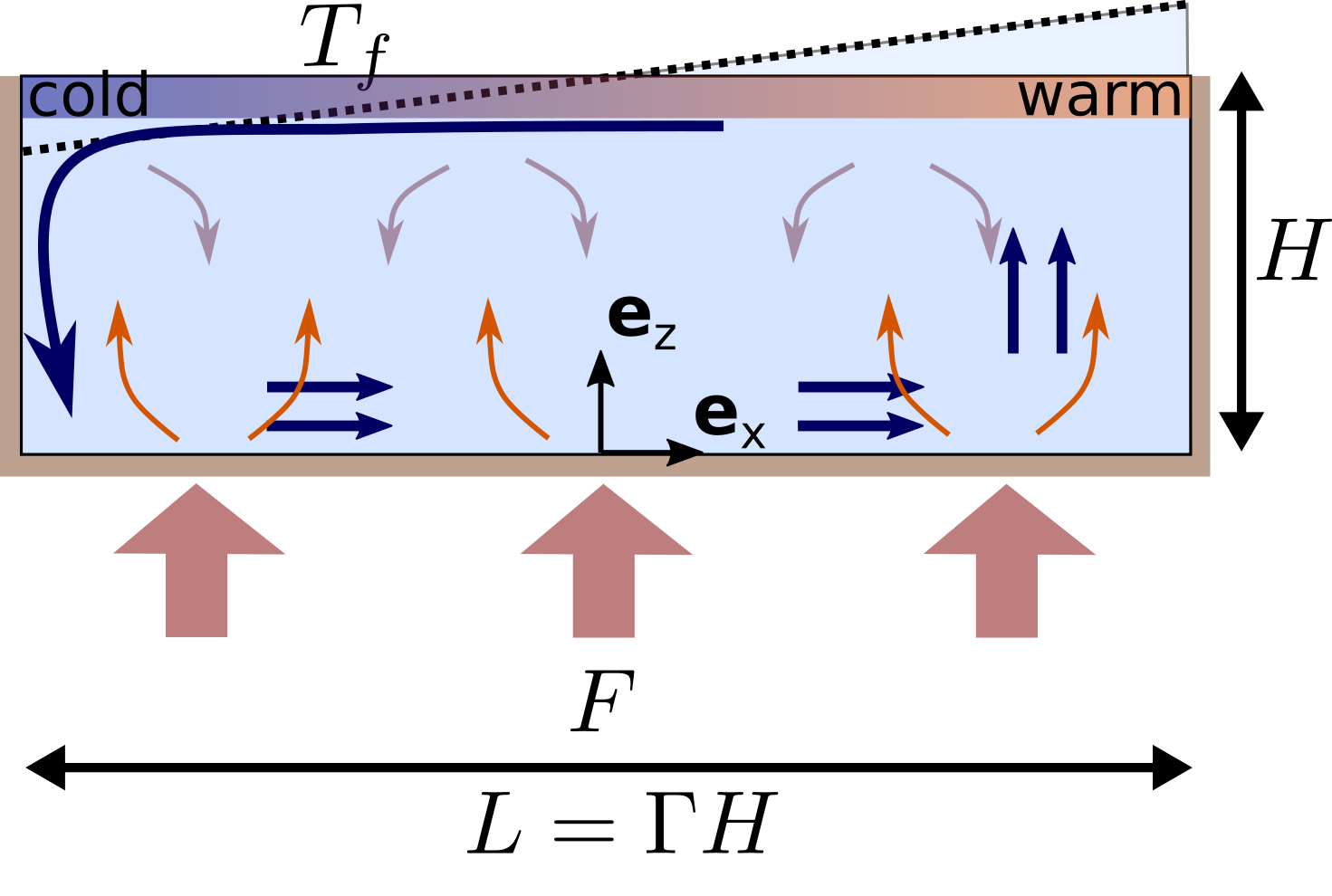}
\includegraphics[width=0.45\textwidth]{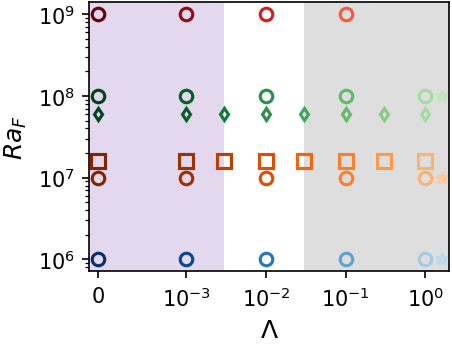}
\put(-350,115){\large{(a)}}
\put(-180,130){\large{(b)}}
\vspace{-0.1in}\caption{(a) Problem schematic. A rectangular fluid domain is subject to bottom heating and a horizontal temperature gradient along the top boundary. The former generates warm rising plumes contributing to multiple overturning cells (thin arrows), while the latter drives a single cell with intense down-welling near and below the cold end of the top boundary (thick arrow). In a subglacial lake, a tilted ice-water interface (shown by the dotted line), due to variable ice thickness above the lake water, would drive horizontal convection because of the pressure-dependence of the freezing temperature. (b) Phase diagram of this study. {The circles show the location of the simulations in $(\Lambda,Ra_F)$ space for $\Gamma=8$; for $\Gamma=4$ (resp. $\Gamma=16$) the markers are diamonds (squares) and are slightly shifted downward (upward) with respect to $Ra_F=10^8$ ($Ra_F=10^7$) for better visibility. The stars show pure HC simulations run with $\Lambda=1$ (with markers shifted to the right for better visibility).} The purple and grey shadings highlight regions of the parameter space wherein the regime dynamics is similar to {Rayleigh--Bénard} convection and horizontal convection, respectively.}
\label{fig1}
\end{figure}

We consider a two-dimensional rectangular fluid domain with Cartesian coordinates $(x,z)$ centred on the bottom boundary; $\bs{e}_z$ is the upward-pointing unit vector of the $z$ axis, which is opposite to gravity (figure \ref{fig1}(a)). We denote by $H$ and $L$ the fluid depth and domain length. {We consider a pure fluid (similar to fresh water), i.e. without dissolved salts.} The evolution of the fluid velocity $\u$, pressure $p$ and temperature $T$ are governed by the {Navier--Stokes} {equations} in the Boussinesq approximation, i.e.
\bsa{eq:a1}\label{eq:a11}
& \p_t \u - \nu\DEL\u + \NA (p/\rho_0) = -  \lp\u\cdot\NA\rp\u    - (\rho'/\rho_0) g\bold{e}_z, \\ \label{eq:a12}
&\NA\cdot \u = 0,  \\ \label{eq:a13}
& \p_t T - \kappa \DEL T = - \lp\u\cdot\NA\rp T,
\esa
where $\rho_0$ is the reference fluid density and $\rho'$ is the density anomaly; $g$ is surface gravity, $\p_t$ denotes time derivative and $\NA$ is the gradient operator. We consider constant thermodynamic and transport parameters, which is known as the Oberbeck approximation, such that the equation of state is simply $\rho'=-\rho_0\alpha (T-T_0)$ with $T_0$ the reference temperature and with the thermal expansion coefficient $\alpha$, along with {kinematic} viscosity $\nu$ and thermal diffusivity $\kappa$, taken constant. All boundaries are no slip. We impose a uniform heat flux on the bottom boundary, a variable temperature profile along the top plate and adiabatic side walls, such that the boundary conditions read
\bsa{eq:bcs}
&\u(z=0)=\u(z=H)=\u\lp x=\pm L/2\rp=\bold{0}, \\ 
&\p_zT(z=0)=-\f{F}{k}, \quad T(z=H)=T_f(x)=\f{\lambda L}{2}\sin\lp\f{\pi x}{L}\rp, \quad \p_xT\lp x=\pm L/2\rp=0.
\esa
It may be noted that the sinusoidal profile of temperature imposed on the top boundary differs from the step-wise or linear profile used in previous studies of horizontal convection, the latter being consistent at leading order with a constant-tilt ice-water interface. We chose a sinusoidal profile because it is consistent with the adiabatic vertical walls on the sides. We use subscript $_f$ to denote the top temperature $T_f$, because the top of the water column must be at the freezing temperature in a subglacial lake. For simplicity, we consider a flat horizontal top boundary. {Note that we often refer to the heat flux imposed on the bottom boundary as the geothermal flux as our study is motivated by subglacial lakes.}  

In order to identify the minimum number of independent parameters and explore their effect on the fluid dynamics, we non-dimensionalize the governing equations \eqref{eq:a1} and boundary conditions \eqref{eq:bcs}. We use the water depth $H$ as characteristic length scale, the diffusive time $\tau_{\kappa}=H^2/\kappa$ as time scale, the velocity {$u_{\kappa}=H/\tau_{\kappa}$} as velocity scale, the temperature difference due to geothermal heating $\Delta = FH/k$ as temperature scale, and the pressure $p_{\kappa}=\rho_0u_{\kappa}^2$ as pressure scale. Using $T_0$ and $p_0+\rho_0g(H-z)$ as temperature and pressure gauges ($p_0$ is the reference pressure), such that we remove the leading-order mean buoyancy and hydrostatic pressure terms that balance each other, we then define dimensionless variables (denoted by tildes) as 
\ba{}\label{eq:02}
(x,z)= H(\widetilde{x},\widetilde{z}), \; t=\tau_{\kappa}\widetilde{t}, \; u= u_{\kappa}\widetilde{u}, \; p = p_0+\rho_0 g(H-z) + p_{\tau}\widetilde{p}, \; T = T_0 + {\Delta} \widetilde{T}. 
\ea
Substituting \eqref{eq:02} into \eqref{eq:a1} and \eqref{eq:bcs} yields a set of dimensionless equations and boundary conditions, which, dropping tildes, read
\bsa{eq:a2}\label{eq:a21}
& \p_{t} \u - Pr\nabla^2\u + \NA p = -  \lp\u\cdot\NA\rp\u + PrRa_FT\bold{e}_z, \\ \label{eq:a22}
& \NA\cdot \u = 0, \\ \label{eq:a23}
& \p_{t} T - \nabla^2 T = - \lp\u\cdot\NA\rp T, 
\esa
and
\bsa{eq:bcs2}
&\u(z=0)=\u(z=1)=\u(x=\pm \Gamma/2)=\bold{0}, \\ 
&\p_zT(z=0)=-1, \quad T(z=1)=\f{\Lambda \Gamma}{2}\sin\lp\f{\pi x}{\Gamma}\rp, \quad \p_xT(x=\pm \Gamma/2)=0,
\esa
wherein{
\ba{} 
Pr\equiv\f{\nu}{\kappa}, \quad Ra_F\equiv\f{\alpha g H^4 F}{k\nu\kappa}, \quad \Lambda\equiv\f{\lambda k}{F}, \quad \Gamma\equiv\f{L}{H},
\ea
}are the control parameters with $Pr$ the Prandtl number, $Ra_F$ the flux-based Rayleigh number, $\Lambda$ the flux ratio and $\Gamma$ the aspect ratio. {Horizontal convection studies typically use a horizontal Rayleigh number as a proxy for the strength of the buoyancy anomaly along the top (or bottom, as the case may be) boundary, which is defined with our notation as
\ba{}
Ra_L\equiv Ra_F\Lambda\Gamma^4.
\ea 
}In this study, classical RB simulations are run with $\Lambda=0$, while classical HC simulations are run with $\Lambda> 0$ (typically $\Lambda=1$) and $Ra_F> 0$ (such that $Ra_L> 0$) but with an insulating bottom boundary, i.e. such that the temperature condition \eqref{eq:bcs2} for $z=0$ is replaced with $\p_z T(z=0)=0$. {Note that while $Ra_L$ is a commonly employed Rayleigh number in HC studies, a temperature-based Rayleigh number, typically written as $Ra_{\Delta}$, is usually preferred over the flux-based Rayleigh number $Ra_F$ in RB convection studies (primarily because RB convection studies typically consider fixed-temperature boundary conditions). Here we will most often use $Ra_F$ as control parameter to describe RB results, but use $Ra_{\Delta}$, which is related to $Ra_F$ through $Ra_{\Delta}=Ra_F/Nu_{RB}$ \citep{Johnston2009} with $Nu_{RB}$ the RB-specific Nusselt number (defined in \S~\ref{sec:nuss}), when appropriate (e.g. in Appendix \S~\ref{appC}).}

We solve equations \eqref{eq:a2} with the open-source spectral-element code Nek5000 \citep{Fischer1997,Deville2002}, which has been extensively used recently in thermal convection studies \citep{Scheel2013,Leard2020}.
The governing equations are cast into weak form and discretised in space by the Galerkin approximation.
The Cartesian domain is discretised using $n_z$ elements in the vertical direction and $\Gamma n_z$ elements in the horizontal direction. Elements have been refined close to all boundaries to properly resolve viscous and thermal boundary layers.
The velocity is discretised within each element using Lagrange polynomial interpolants based on tensor-product arrays of Gauss–Lobatto–Legendre quadrature points.
The polynomial order $l_d$ of the expansion basis on each element varies between $7$ and $11$ in this study. We use the 3/2 rule for dealiasing, i.e. with extended dealiased polynomial order $3/2l_d$.
Convergence has been tested by gradually increasing the polynomial order for a fixed number of elements \citep{Scheel2013}.
The nonlinear terms are treated explicitly by a second-order extrapolation scheme whereas the viscous terms are treated implicitly by a second-order backward differentiation scheme.
The list of simulations performed with corresponding physical and numerical parameters is provided in table \ref{tab:sims} in Appendix \ref{appA}. 

The initial state is motionless and has uniform temperature distribution superimposed with low-amplitude background noise (except in section \S~\ref{sec:hyst} where we perform continuation).
We explore the fluid dynamics in the $(Ra_F,\Lambda)$ parameter space as shown in figure \ref{fig1}(b), i.e. with $10^6 \leq Ra_F \leq 10^9$ and $0 \leq \Lambda \leq 1$. Simulations with $\Lambda=0$ yield canonical {Rayleigh--Bénard} results; for completeness we also run simulations without geothermal flux, i.e. setting $\p_z T =0$ at $z=0$, which yield canonical horizontal convection results. For simplicity, we set $Pr=1$ in all simulations. Most results are shown for $\Gamma=8$, though we also ran simulations with $\Gamma=4$ and 16 to partially assess the effect of the aspect ratio{; $4\leq\Gamma\leq 16$ spans enclosure aspect ratios that are commonly used in HC studies (either experimental or numerical), including $\Gamma = 6.2$ \citep{Mullarney2004,Gayen2013,Gayen2014,Tsai2020} and $\Gamma = 10$ \citep[e.g.][]{Shishkina2016b}.} Note that different colors highlight different $Ra_F$ in figure \ref{fig1}(b), whereas squares, circles and diamonds denote results obtained for $\Gamma=4,$ 8 and 16, respectively; stars show the results of pure (no geothermal flux) horizontal convection simulations. As we will show, $\Lambda$ is a better indicator of the regime dynamics than $Ra_L/Ra_F$. The range of $0\leq \Lambda\leq 1$ values is motivated by subglacial lakes, as explained in Appendix \ref{appB}. 

\section{Results}\label{sec:res}

\subsection{Flow regimes}\label{sec:phen}

\begin{figure}
\centering
\includegraphics[width=1\textwidth]{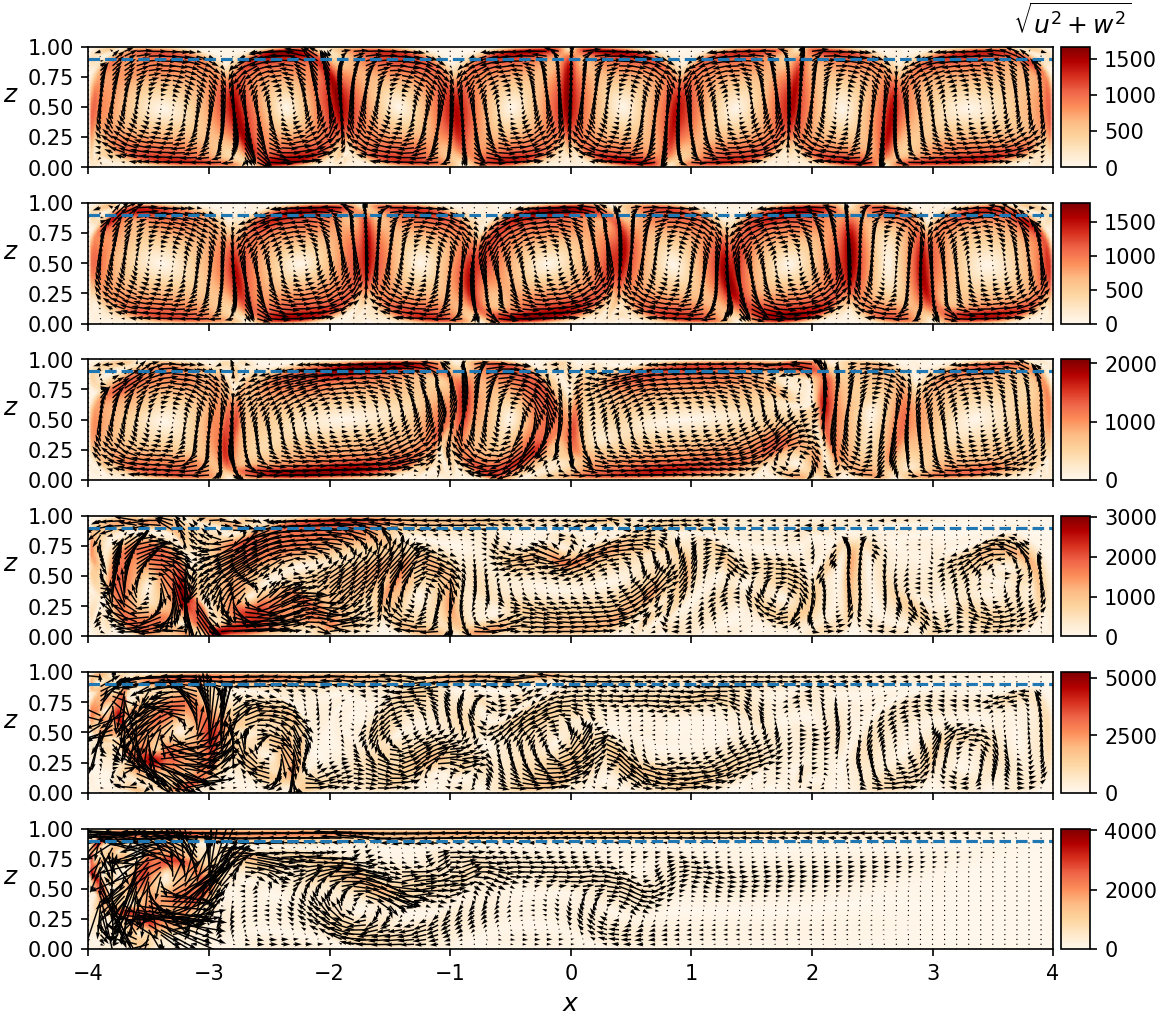}
\put(-210,324){{$\Lambda=0$}}
\put(-210,272){{$\Lambda=10^{-3}$}}
\put(-210,220){{$\Lambda=10^{-2}$}}
\put(-210,168){{$\Lambda=10^{-1}$}}
\put(-210,117){{$\Lambda=1$}}
\put(-280,65){{Pure HC (no geothermal flux) with $Ra_L\approx 4\times 10^{11}$}}
\vspace{-0.1in}\caption{Snapshots of the velocity magnitude $\sqrt{u^2+w^2}$ and velocity vector field (shown by arrows) at the end of the simulations with $Ra_F=10^8$ and $\Lambda$ increasing from top to bottom. The bottom figure shows the results obtained for pure HC, i.e. with $\Lambda=1$ but with an insulating bottom boundary (no geothermal flux). The characteristic length scale of overturning motions is estimated from the horizontal flow at $z=0.9$ (see \S~\ref{sec:auto}), which is highlighted by the blue dashed lines.}
\label{fig2}
\end{figure}
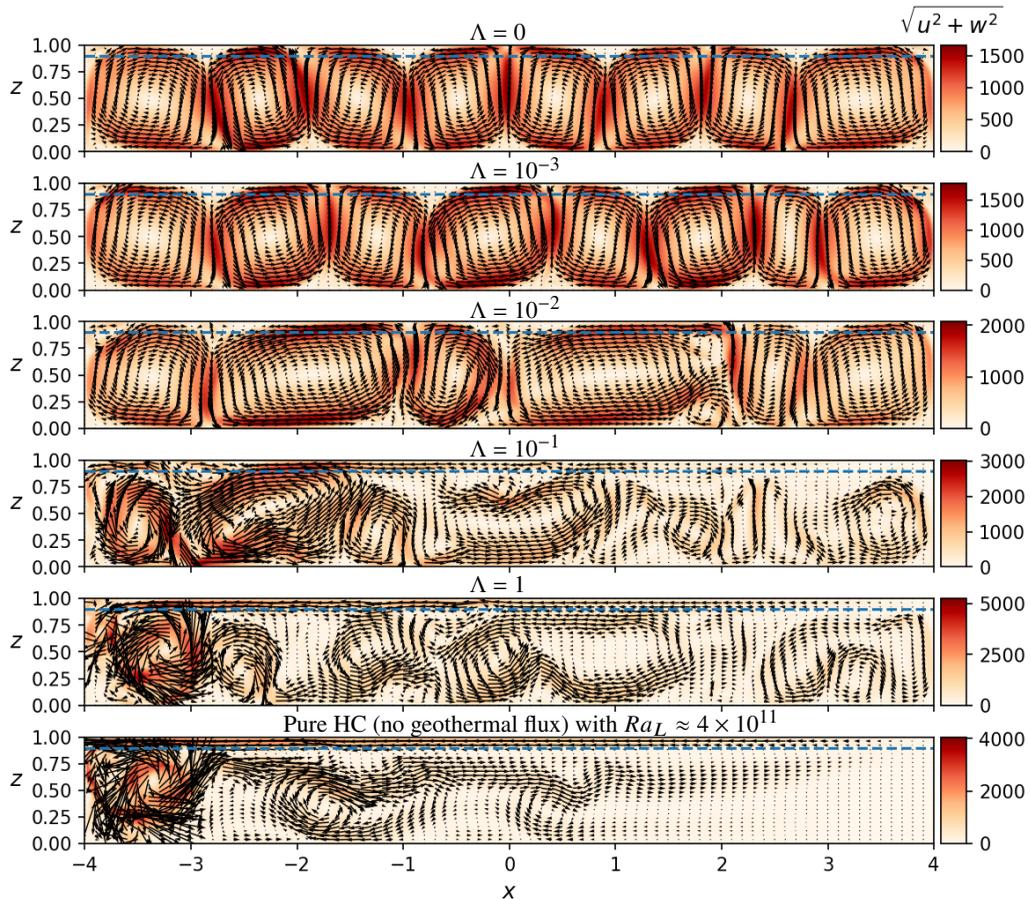

\begin{figure}
\centering
\includegraphics[width=1\textwidth]{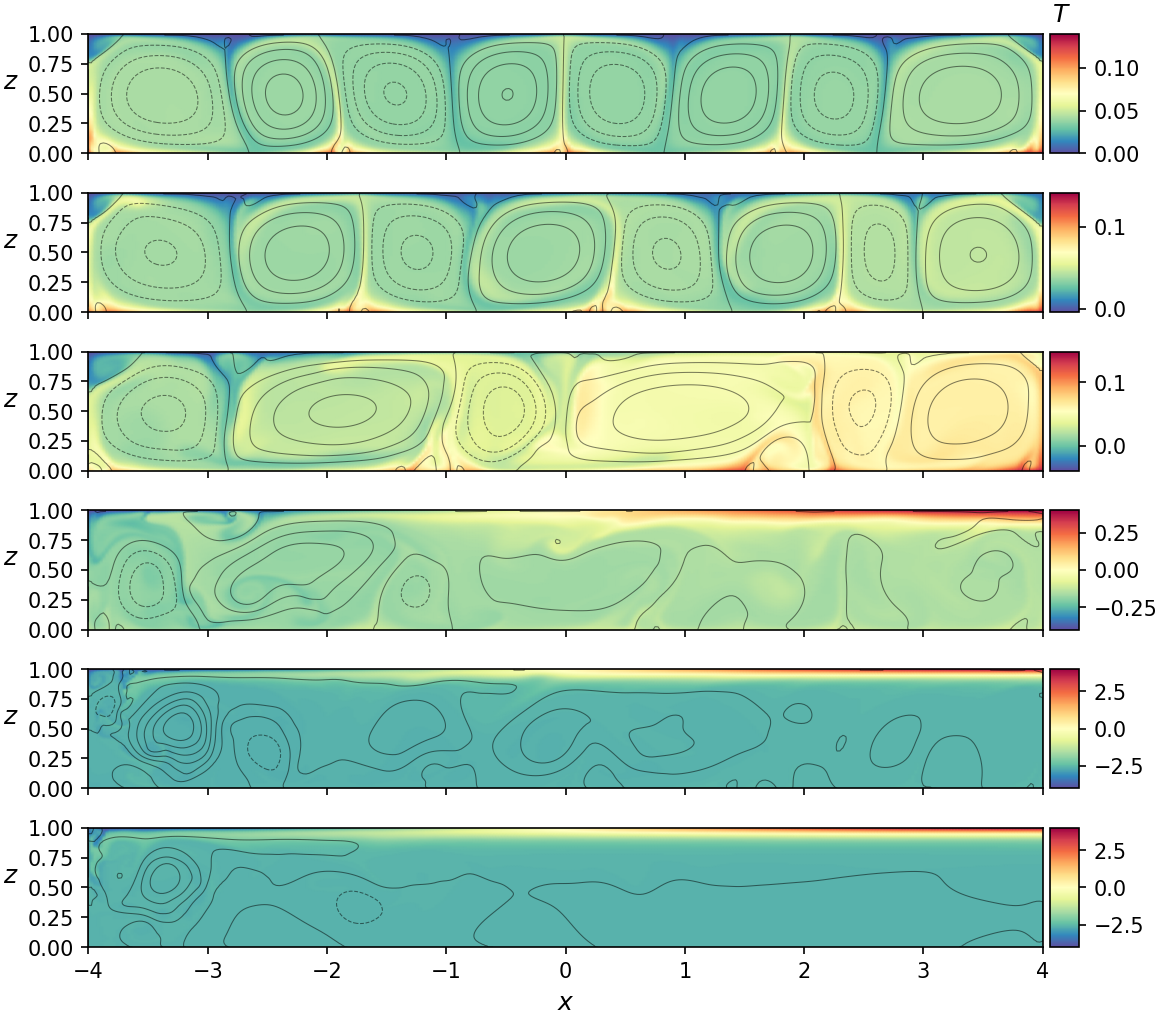}
\put(-210,327){{$\Lambda=0$}}
\put(-210,275){{$\Lambda=10^{-3}$}}
\put(-210,222){{$\Lambda=10^{-2}$}}
\put(-210,170){{$\Lambda=10^{-1}$}}
\put(-210,118){{$\Lambda=1$}}
\put(-280,65){{Pure HC (no geothermal flux) with $Ra_L\approx 4\times 10^{11}$}}
\vspace{-0.1in}\caption{Snapshots of the temperature field at the end of the simulations with $Ra_F=10^8$ and $\Lambda$ increasing from top to bottom. {The solid (resp. dashed) lines show positive (resp. negative) contours of the streamfunction, which is set to 0 at the bottom left corner.} The bottom figure shows the results obtained without geothermal heating. Note that the color bars are not the same between plots.}
\label{fig3}
\end{figure}

We first show in figures \ref{fig2} and \ref{fig3} snapshots of the velocity and temperature fields at statistical steady state for $Ra_F=10^8$ and $\Gamma=8$ with $\Lambda$ increasing from top to bottom.
The top plot in figure \ref{fig2} shows the velocity field obtained for $\Lambda=0$, which is the canonical RB case.
The flow is organized in 4 pairs of counter-rotating rolls, with characteristic length scale approximately equal to twice the domain height, as expected from linear stability analysis \citep{Chandrasekhar1961}.
As $\Lambda$ increases, the overturning cells become distorted because of the buoyancy anomaly on the top boundary, which triggers preferentially leftward flows in the upper half of the domain.
For $\Lambda=10^{-2}$ (3rd plot from the top in figure \ref{fig2}), only 3 pairs of counter-rotating cells co-exist, and the 3 counter-clockwise cells are elongated because the upper leftward flow branch they support is enhanced by the top boundary.
As $\Lambda$ increases above $10^{-2}$, the flow becomes dominated by horizontal convection, which includes intense down-welling below the cold end (left) of the top boundary that trigger vigorous local overturning, and a large-scale counter-clockwise current, which inhibits the growth of RB-like cells in the rest of the domain.
The velocity fields obtained for $\Lambda=1$ with (2nd plot from bottom) and without (bottom plot) bottom heat flux are similar, although the former displays stronger meanders of the large-scale flow near the bottom boundary.

The temperature field in figure \ref{fig3} also highlights the existence of RB-like cells for small $\Lambda$, which merge at intermediate $\Lambda$ values.
For $\Lambda=1$ (2 bottom plots of figure \ref{fig3}), the bulk temperature becomes significantly smaller than 0, which is the average temperature of the top boundary, because mixing occurs preferentially on the left of the domain where the top fluid is cold and the down-welling is intense; a warm layer develops near the top right-hand side of the domain but does not mix with the bulk as it is locally stably stratified.
It is noteworthy to remark that the bulk temperature is cooler on the left-hand side than on the right-hand side of the domain at intermediate $\Lambda=10^{-2}$ (3rd plot from the top). This horizontal temperature gradient of the bulk, which is maintained across multiple cells, builds up until it becomes so great that a single counter-rotating large-scale flow takes over. However, the large-scale flow is short lived because $\Lambda$ is small, such that another cycle of RB cells emerges until the horizontal temperature gradient becomes too large again. This subtle bursting dynamics points toward the possible existence of hysteresis, which we investigate in \S~\ref{sec:hyst}.     

\subsection{Mean temperature and Reynolds number}\label{sec:numb}

In sections \S~\ref{sec:numb} and \S~\ref{sec:nuss}  we first show that all simulations reach a statistical steady state. {Then we explore the scaling trends of the Reynolds and Nusselt numbers with the problem parameters, and we demonstrate that the latter can be used to distinguish RB- from HC-dominated simulations}. We use $\langle X \rangle_x$ to denote $x$-averaged variables, $\langle X \rangle$ to denote volume-averaged variables, and over-line $\overline{X}$ to denote temporal averages at statistical steady state (typically from $t\geq 1$ onward). Whenever relevant, we show the standard deviation (due to temporal fluctuations) of averaged quantities with vertical error bars. Note, however, that the standard deviation is always small, such that the error bars are often smaller than the marker size and thus barely visible. 

\begin{figure}
\centering
\includegraphics[width=0.92\textwidth]{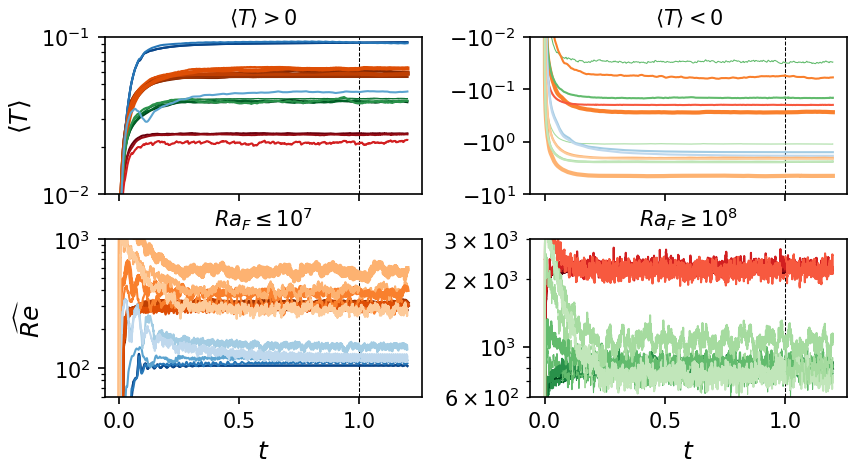}
\includegraphics[width=0.07\textwidth]{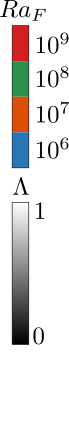}
\put(-379,185){\large{(a)}}
\put(-209,185){\large{(b)}}
\put(-379,102){\large{(c)}}
\put(-209,102){\large{(d)}}
\vspace{-0.1in}\caption{Time history of volume-averaged (top row) temperature and (bottom row) Reynolds number. {Simulation results are split into two different subplots (left and right columns) for better visibility (as indicated by the plot titles)}. Different colors correspond to different Rayleigh numbers $Ra_F$, lighter colors indicate higher $\Lambda$, and line width increases with $\Gamma$. Time-averaged variables representative of the statistical steady state are integrated from $t=1$ (shown by the vertical dashed lines) onward.}
\label{fig4}
\end{figure}

\begin{figure}
\centering
\includegraphics[width=0.8\textwidth]{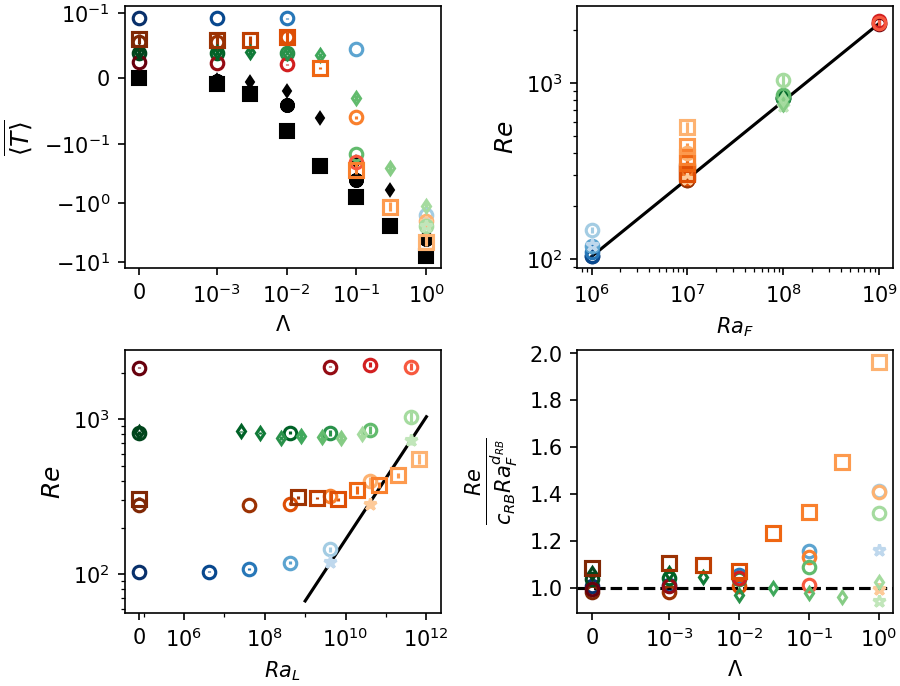}
\includegraphics[width=0.12\textwidth]{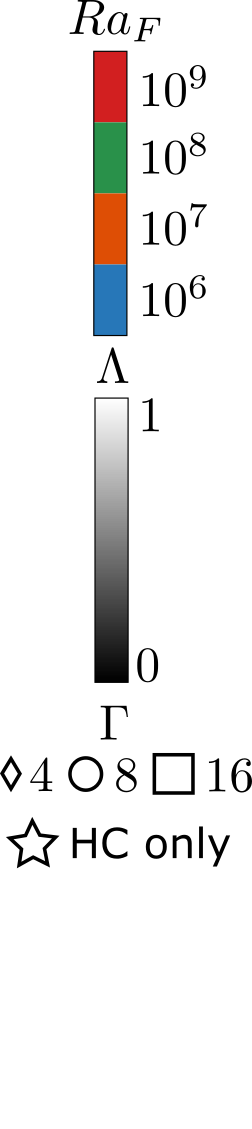}
\put(-350,233){\large{(a)}}
\put(-193,233){\large{(b)}}
\put(-138,147){\rotatebox{40}{\scriptsize{$d_{RB}=0.441$}}}
\put(-350,113){\large{(c)}}
\put(-248,24){\rotatebox{55}{\scriptsize{$d_{HC}=0.396$}}}
\put(-193,113){\large{(d)}}
\vspace{-0.1in}\caption{(a) Mean temperature at statistical steady state for all simulations as a function of the flux ratio $\Lambda$. The black symbols show the coldest temperature on the top boundary, i.e. $T=-\Lambda\Gamma/2$. (b),(c) Reynolds number $Re$ as a function of $Ra_F$ and $Ra_L$, respectively, with scalings $Re=c_{RB}Ra_F^{d_{RB}}$ and $Re=c_{HC}Ra_L^{d_{HC}}$ shown by the black solid lines. (d) Compensated Reynolds number as a function of $\Lambda$. The star symbols show results obtained for pure HC simulations.}
\label{fig5}
\end{figure}
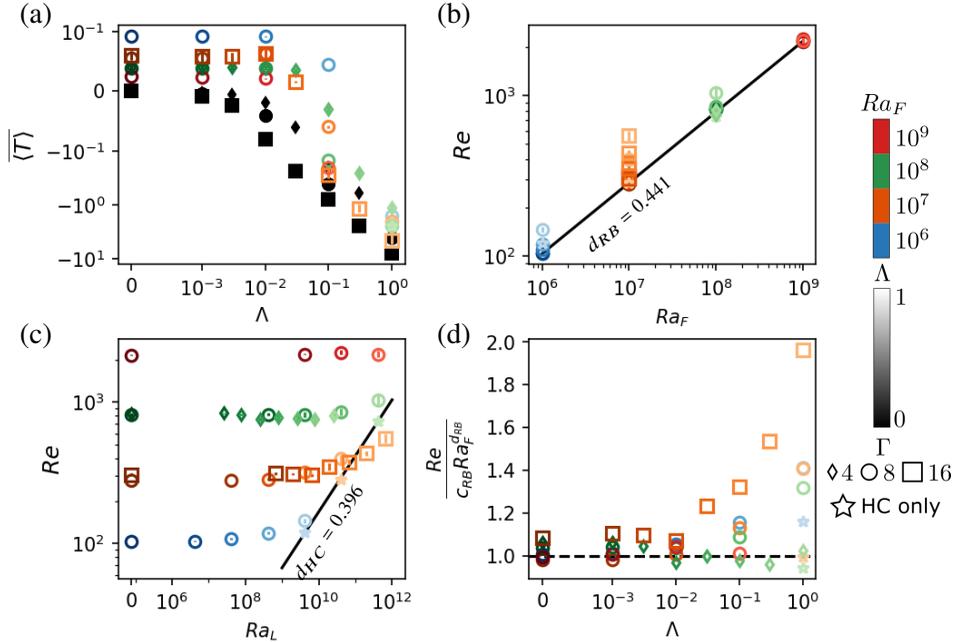

Figure \ref{fig4} shows the temporal evolution of the volume-averaged temperature $\langle T \rangle$ (top row) and Reynolds number $\widehat{Re}$ (bottom row). In this study, we use the kinetic energy density to construct the Reynolds number, i.e. such that
\ba{}
\widehat{Re} \equiv \sqrt{\langle u^2 + w^2 \rangle}, 
\ea
which is close to the Reynolds number based on the velocity root mean square (not shown). The time-averaged Reynolds number is
\ba{}
Re \equiv \sqrt{\overline{\langle u^2 + w^2 \rangle}}.
\ea
Both $\langle T \rangle$ and $\widehat{Re}$ display a sharp transient followed by a statistical steady state (small fluctuations around a mean value independent of time) at approximately $t=0.5$; thus, here we use (conservatively) $t=1$ as the initial time for time averaging of output variables representative of the statistical steady state. 
For small $\Lambda$ (figure \ref{fig4}(a)), the mean temperature increases from 0 up to a small but positive value as a result of geothermal heating, which dominates over horizontal convection. For large $\Lambda$ (figure \ref{fig4}(b)), down-welling beneath the cold end of the top boundary cools down the fluid more efficiently than geothermal heating can warm it, such that the bulk temperature becomes {negative}. The Reynolds number exhibits stronger fluctuations in time than the mean temperature, which are on the overturning time scale. As expected, $\widehat{Re}$ increases with $Ra_F$, as can be seen from the stacks of lines shown by blue ($Ra_F=10^6$), orange ($Ra_F=10^7$), green ($Ra_F=10^8$) and red ($Ra_F=10^9$) colors that are successively on top of each other (results shown in figure \ref{fig4}(d) would be above those shown in figure \ref{fig4}(c)). 
{The effect of $\Lambda$ (color intensity) and $\Gamma$ (line thickness) on $\widehat{Re}$, which is relatively weak for our set of simulations (although clearly visible for $Ra_F=10^7$ and $Ra_F=10^8$ in figure \ref{fig4}), is commented on in greater details when discussing figure \ref{fig5}(d).}  

We show in figure \ref{fig5} the time-averaged mean temperature $\overline{\langle T \rangle}$ and Reynolds number $Re$ at statistical steady state as functions of the problem parameters. Figure \ref{fig5}(a) shows that the mean temperature decreases quickly with increasing $\Lambda\geq 10^{-2}$ because down-welling below the cold top boundary ($T(x=-\Gamma/2)=-\Lambda\Gamma/2$ shown by black markers) becomes sufficiently strong to lower the bulk temperature. The mean temperature also decreases with increasing $Ra_F$ (blue markers above orange, green and red markers) because increasing geothermal heating makes mixing more efficient, thus lowering temperature differences, while increasing horizontal convection increases mixing from the cold region of the top boundary. 

{Figures \ref{fig5}(b) shows that the power law curve $Re=c_{RB}Ra_F^{d_{RB}}$ (black solid line) provides a good prediction for the Reynolds number as a function of $Ra_F$ for most simulations. Here, pre-factor $c_{RB}$ and exponent $d_{RB}$ are obtained from best fit with the results for $\Lambda=0$ and $\Gamma=8$ (see table \ref{tab:fit} and Appendix \ref{appC} for a list and discussion of all pre-factors and exponents mentioned in the paper). 
The dependence of $Re(Ra_F)$ with $\Lambda$ (and $\Gamma$) is small, especially at low $\Lambda<10^{-2}$ (dark colors), which means that a small horizontal temperature gradient (or change of aspect ratio) has limited effect on the intensity of RB-driven flows. Conversely, figure \ref{fig5}(c) demonstrates that there is a wide spread of $Re(Ra_L)$ between simulations, even for relatively large $\Lambda \sim 0.1$ (light colors), which means that the circulation is almost always affected by the bottom heat flux (if not driven by it), even in the HC-dominated regime obtained for $\Lambda>O(10^{-2})$ (as we will show). Accordingly, the power law curve $Re= c_{HC}Ra_L^{d_{HC}}$ (black solid line) obtained from pure HC results (shown by the stars) predicts $Re$ accurately for large $\Lambda\geq 10^{-1}$ only. }

We plot the Reynolds number compensated by the RB scaling in figure \ref{fig5}(d) in order to highlight the effect of $\Lambda$ and $\Gamma$ on $Re$. {The spread of $Re$ with $\Gamma$ and $\Lambda$ is overall small (less than 40\% increase) except for $Ra_F=10^7$ (up to 100\% increase). At low $\Lambda$, the increase of $Re$ with $\Gamma$ is modest (about 10\% or less), in agreement with previous two-dimensional RB studies that have shown a monotonic increase of $Re$ with $\Gamma$ in two-dimensional bounded domains with a $\sim$20\%-increase plateau reached at $\Gamma \approx 10$ \citep{VanDerPoel2012}. At large $\Lambda$, the increase of $Re$ with $\Gamma$ becomes more important (up to 50\%) as HC, which is controlled by $Ra_L\propto \Gamma^4$, starts dominating. An increase of $Re$ with $\Lambda$ may be expected at large $\Lambda$, i.e. once HC dominates, since $Ra_L\propto \Lambda$. Here, the increase of $Re$ with $\Lambda$ is significant only for the set of simulations with $Ra_F=10^7$ and $\Gamma=16$, i.e. which attain the largest $Ra_L$ (figure \ref{fig5}(c)). This suggests that our simulations transition from RB to HC (around $\Lambda\approx 10^{-2}$, as will be shown) primarily through a change of flow structure, not intensity ($Re$ almost fixed).} The exact scaling of $Re$ with $\Lambda$ and $\Gamma$ in the HC regime is beyond the scope of our study given the limited amount of data in the large $\Lambda$ and $\Gamma$ limit. Note, however, that the enhancement of $Re$ by geothermal heating in the HC regime seen in figure \ref{fig5}(d)  (compare star symbols and circles at large $\Lambda$) is compatible with previous studies \citep{Mullarney2006,Wang2016}. {Both studies indeed reported a $O(1)$ increase due to geothermal heating in the volume flux (measured through the maximum of the streamfunction), which may be expected to scale linearly with $Re$, for $\Lambda Nu^{\chi}_{HC}\sim O(10)$, which is equivalent to $\Lambda \sim O(0.1)$--$O(1)$ as the Nusselt number of HC (defined in section \S~\ref{sec:nuss}) $Nu^{\chi}_{HC}\sim O(10)$--$O(100)$ for $Ra_L\sim O(10^{9})$--$O(10^{12})$ \citep{Mullarney2004}.}

\begin{table}\centering 
\begin{tabular}{C{4cm}C{2cm}C{1.5cm}}
\hline
power law & pre-factor & exponent  \\ 
$Re=c_{RB}Ra_F^{d_{RB}}$ & $2.326\times 10^{-1}$ & $0.441$ \\
$Re=c_{HC}Ra_L^{d_{HC}}$ & $1.839\times 10^{-2}$ & $0.396$ \\
$Nu_{RB}=a_{RB}Ra_F^{b_{RB}}$ & $3.938\times 10^{-1}$ & $0.191$ \\
$Nu^{\chi}_{HC}=a_{HC}Ra_L^{b_{HC}}$ & $3.181\times 10^{-1}$ & $0.226$ \\
\hline
\end{tabular}\vspace{-0.in}\caption{Pre-factor and exponent of all power laws referred to in the paper and shown as black solid lines in figures \ref{fig5}(b),(c), \ref{fig7a}(a) and \ref{fig7b}(a). The power laws for $Re(Ra_F)$ and $Nu(Ra_F)$ are based on four simulations with $\Lambda=0$ and $\Gamma=8$, whereas the power laws for $Re(Ra_L)$ and $Nu(Ra_L)$ are based on three simulations without geothermal flux and $\Gamma=8$ (see table \ref{tab:sims} in Appendix \ref{appA}).}\label{tab:fit}\end{table}

\subsection{Heat transfer efficiency via Nusselt numbers}\label{sec:nuss}

In order to assess the efficiency of heat transfer, we define two distinct Nusselt numbers designed to measure heat transfer due to either {Rayleigh--Bénard} or horizontal convection. The Nusselt number of RB convection is defined as
\ba{}\label{eq:nurb} 
Nu_{RB} \equiv \f{FH}{k\lb\overline{\langle T^{\mathrm{dim}}(z=0) \rangle_x}-\text{min}(T^{\mathrm{dim}})\rb} = \f{1}{\overline{\langle T(z=0) \rangle_x}-\text{min}(T)},
\ea
where superscript $^\mathrm{dim}$ denotes a dimensional variable and $\text{min}(T)=-\Lambda\Gamma/2$ is the coldest temperature achieved on the top boundary. As is customary in classical RB studies, $Nu_{RB}$ compares the effective vertical heat flux that comes out of the system to the vertical heat flux that would be obtained from conduction only due to the temperature difference between the top and bottom boundaries. Here we use $\text{min}(T)$ instead of the mean $T=0$ value as a gauge for the top temperature in the denominator in equation \eqref{eq:nurb} to ensure $Nu_{RB}>0$, since $\overline{\langle T(z=0) \rangle_x}$ becomes negative for large-enough $\Lambda$ (figure \ref{fig5}(a)). {We note that the use of $\min(T)$ affects the definition of the heat flux of the diffusive state; however, in the RB regime, i.e. for $\Lambda\ll 10^{-2}$, $\min(T)$ is always much smaller than the mean bottom temperature, such that $Nu_{RB}$ does converge toward the commonly-defined Nusselt number of RB convection.}

\begin{figure}
\centering
\includegraphics[width=0.75\textwidth]{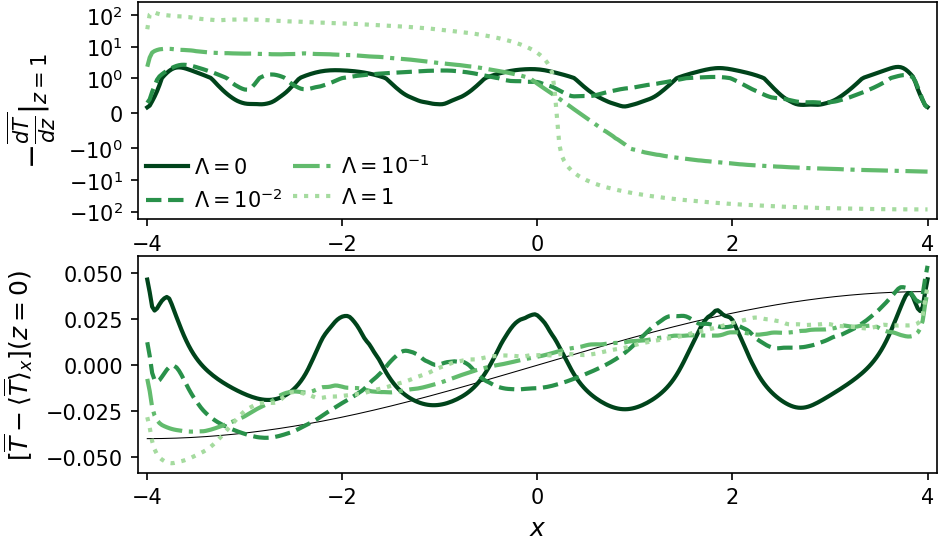}
\put(-300,165){\large{(a)}}
\put(-300,87){\large{(b)}}
\vspace{-0.1in}\caption{(a) Time-averaged heat flux along the top boundary for $Ra_F=10^8$, $\Gamma=8$ and four different $\Lambda$ values (see legend). {(b) Same as (a) but for the temperature on the bottom boundary with the horizontal mean removed. The thin black solid line shows the temperature on the top boundary for $\Lambda=10^{-2}$.}}
\label{fig6}
\end{figure}

Unlike RB convection, HC sets up a large-scale circulation, which produces large asymmetry between the left-hand and right-hand sides of the fluid domain: heat extraction occurs below the cold (left) end of the top boundary while heat is replenished on the warm (right) end via a slow return flow and a thick boundary layer. This asymmetry can be seen in figure \ref{fig6}(a), which shows the time-averaged heat flux on the top boundary for RB-dominated, mixed RB-HC and HC-dominated simulations with $Ra_F=10^8$: the RB results ($\Lambda=0$) show an oscillatory pattern of heat flux linked to the underlying overturning cells, the intermediate $\Lambda=10^{-2}$ value leads to skewed oscillations and the large $\Lambda=1$ value yields a much-larger monotonic and anti-symmetric heat flux pattern with respect to the middle of the domain $x=0$. {Figure \ref{fig6}(b) shows similar patterns for the temperature on the bottom boundary. The temperature increase from left to right imposed on the top boundary is imprinted on the bottom boundary quite clearly for large $\Lambda$ but also for $\Lambda$ as small as $10^{-2}$.} 

{Different Nusselt numbers have been used in the past to measure heat transfer by horizontal convection, such as the Nusselt number based on the intensity in absolute value of heat extraction and deposition at the top boundary \citep{Rossby1965,Rossby1998}, i.e. 
\ba{}\label{eq:nuabshc}
Nu^{\text{abs}}_{HC} \equiv \f{\overline{\langle \left|\p_z T (z=1)\right| \rangle_x}}{ \langle \left|\p_z T_{\mathrm{diff}}(z=1)\right| \rangle_x},
\ea
and the Nusselt number based on the heat flux above the cold (or warm) half of the boundary \citep{Sheard2011}, i.e.
\ba{}\label{eq:nucoldhc}
Nu^{\text{half}}_{HC} \equiv \f{\overline{\langle \p_z T (z=1) \rangle_{x<0}} }{ \langle \p_z T_{\mathrm{diff}}(z=1) \rangle_{x<0} },
\ea
where subscript $_{\mathrm{diff}}$ means \textit{of the diffusive base state}.
While $Nu^{\text{abs}}_{HC}$ and $Nu^{\text{half}}_{HC}$ are intuitive, they are not based on fundamental variables of the energy budget.
Thus, following the recommendation by \citet{Rocha2020} we use another definition of the Nusselt number, i.e.
\ba{}
Nu^{\chi}_{HC} \equiv \f{\chi}{\chi_{\mathrm{diff}}},
\ea
}where $\chi=\overline{\langle |\NA T|^2 \rangle}$ is the dissipation of temperature variance, which is related to the horizontal heat transport and entropy production. We show in Appendix \ref{appD} that $\chi$ is proportional to the correlation of the heat flux with the temperature on the bounding horizontal plates (as already shown by \citet{Rocha2020} without geothermal heating), such that the Nusselt number can be calculated from line integrals as
\ba{}\label{eq:nuhc}
Nu^{\chi}_{HC} = \f{\overline{\langle T(z=1)\p_z T(z=1) - T(z=0)\p_z T(z=0) \rangle_x}}{\overline{\langle T_{\mathrm{diff}}(z=1)\p_z T_{\mathrm{diff}}(z=1) - T_{\mathrm{diff}}(z=0)\p_z T_{\mathrm{diff}}(z=0) \rangle_x}},
\ea
with an analytical expression for the denominator given in equation \eqref{eq:budg3}. We would like to note that while $Nu^{\chi}_{HC}=\chi/\chi_{\mathrm{diff}}$ is thermodynamically compelling, {it is very close to $Nu^{\text{abs}}_{HC}$ and $Nu^{\text{half}}_{HC}$ for pure HC simulations. For mixed RBH simulations, differences exist but are due at leading order to the diffusive normalization (see details in Appendix \ref{appE}).} 

\begin{figure}
\centering
\includegraphics[width=0.75\textwidth]{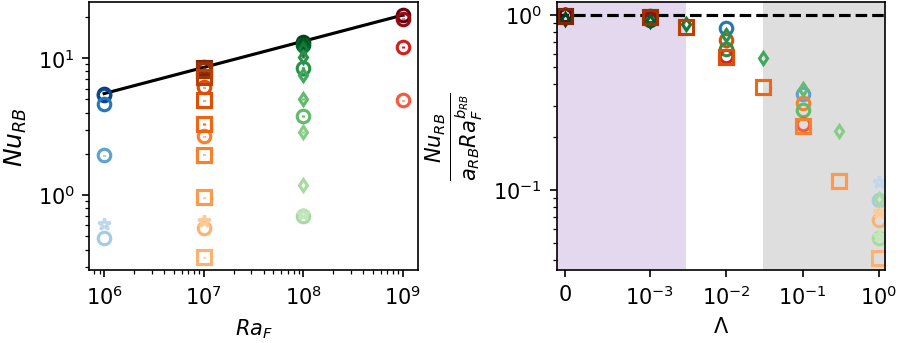}
\includegraphics[width=0.1\textwidth]{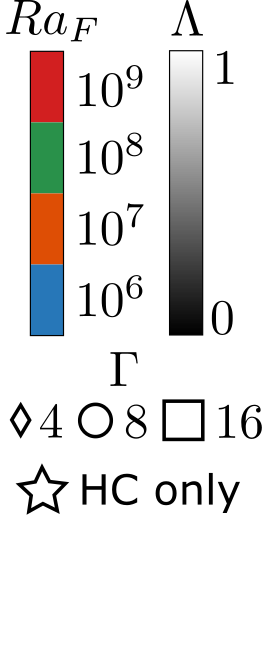}
\put(-323,111){\large{(a)}}
\put(-290,85){\rotatebox{16}{\scriptsize{$b_{RB}=0.191$}}}
\put(-185,111){\large{(b)}}
\vspace{-0.1in}\caption{(a) RB-based Nusselt number $Nu_{RB}$ as a function of $Ra_F$ with scaling law $Nu_{RB}=a_{RB}Ra_F^{b_{RB}}$ shown by the black solid line. (b) Compensated RB-based Nusselt number as a function of $\Lambda$.}
\label{fig7a}
\end{figure}

Figure \ref{fig7a}(a) shows $Nu_{RB}$ (equation \eqref{eq:nurb}) as a function of $Ra_F$.
Simulation results obtained for $\Lambda\leq 10^{-2}$ all collapse very well on the power law curve shown by the black solid line, which was obtained from best fit for $\Lambda=0$ (see table \ref{tab:fit}) and whose exponent is compatible with the classical scaling law of RB convection {for moderate Rayleigh numbers (details in Appendix \ref{appC})}.
The plot of the compensated $Nu_{RB}$ number in figure \ref{fig7a}(b) highlights the deviation of the RB-based Nusselt number from the RB scaling with increasing $\Lambda$. {In particular, the difference between $Nu_{RB}$ and $a_{RB}Ra_F^{b_{RB}}$ becomes of order 1 when $\Lambda \geq 10^{-2}$ (grey area), which we will show marks the transition from RB to horizontal convection.} Note that the decrease of $Nu_{RB}$ with $\Lambda$ is primarily the result of an increase of the denominator in equation \eqref{eq:nurb}, which is due to the fact that the coldest temperature on the top boundary increases more quickly than the mean bottom temperature (in absolute value).

\begin{figure}
\centering
\includegraphics[width=0.89\textwidth]{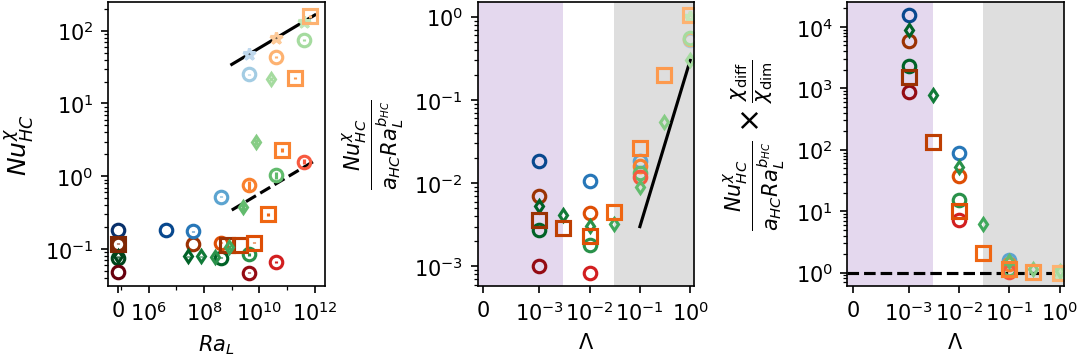}
\includegraphics[width=0.1\textwidth]{grille-de-lecture-des-symboles.png}
\put(-375,113){\large{(a)}}
\put(-258,113){\large{(b)}}
\put(-135,113){\large{(c)}}
\put(-330,87){\rotatebox{28}{\scriptsize{$b_{HC}=0.226$}}}
\put(-175,45){\rotatebox{0}{\scriptsize{$+2$}}}
\vspace{-0.1in}\caption{(a) HC-based Nusselt number $Nu^{\chi}_{HC}$ as a function of $Ra_L$ with scaling law  shown by the black solid line (derived from pure HC simulation results). Note  that the dashed line shows a similar scaling albeit with a different pre factor. (b) Compensated Nusselt number as a function of $\Lambda$. The solid line has a $+2$ slope and shows the scaling of $\chi_{\mathrm{diff}}$ for $\Lambda\ll 1$. (c) Same as (b) but for a Nusselt number with denominator $\chi_{\mathrm{diff}}$ replaced by $\chi_{\mathrm{dim}}=\pi^2\Lambda^2/8$, which discards the effect of geothermal heating and aspect ratio.}
\label{fig7b}
\end{figure}

Figure \ref{fig7b}(a) shows $Nu^{\chi}_{HC}$ as a function of $Ra_L$.
Simulation results obtained for $\Lambda\geq 10^{-2}$ all tend asymptotically (for fixed $\Lambda$, or color intensity) toward power laws parallel (in log-log space) to the one shown by the black solid line (see e.g. the dashed line), which was obtained from best fit for $\Lambda=1$, $\Gamma=8$ and without geothermal flux (cf. perfect overlap with star symbols), and whose exponent is compatible with the  1/5 exponent predicted by \citet{Rossby1965}. {Figure \ref{fig7b}(a) shows that $Nu^{\chi}_{HC}$ can help distinguish simulations dominated by RB convection from simulations dominated by HC: at small $\Lambda$ (dark symbols with typically small $Ra_L$), $Nu^{\chi}_{HC} < 1$, while at large $\Lambda$ (light symbols with typically large $Ra_L$), $Nu^{\chi}_{HC}\gg 1$. The large spread of $Nu^{\chi}_{HC}$ with $\Lambda$ and $\Gamma$ for fixed $Ra_L$ in the HC limit (i.e. at large $\Lambda$) is somewhat unexpected but can be explained: it is due to the diffusive normalization $\chi_{\mathrm{diff}}$. First, figure \ref{fig7b}(b) shows that dividing $Nu^{\chi}_{HC}$ by the pure HC scaling $a_{HC}Ra_L^{b_{HC}}$ results in a steep scaling with $\Lambda$ ($+2$ slope shown by the solid line) for $\Lambda\gg 10^{-2}$, which is exactly the scaling of the diffusive normalization $1/\chi_{\mathrm{diff}} - 1 \propto \Lambda^2$ once Taylor expanded in the small $\Lambda^2$ limit (cf. equation \eqref{eq:budg3}). Second, figure \ref{fig7b}(c) demonstrates that replacing $\chi_{\mathrm{diff}}$ with $\chi_{\mathrm{dim}}=\pi^2\Lambda^2/8$, which is dimensionally-equivalent but discards geothermal heating and approximates $\tanh \lp \pi/\Gamma \rp\approx \pi/\Gamma$ (large $\Gamma$ limit), in the definition of $Nu_{HC}^{\chi}$ (equation \eqref{eq:nuhc}) yields a perfect overlap of all simulation results obtained for large $\Lambda\gg 10^{-2}$ with the power law $a_{HC}Ra_L^{b_{HC}}$. Ultimately, the spread of $Nu_{HC}^{\chi}$ with $\Lambda$ and $\Gamma$ at large $\Lambda$ is due to the diffusive normalization $\chi_{\mathrm{diff}}$ (denominator in \eqref{eq:nuhc}), not $\chi$, because  $\chi_{\mathrm{diff}}$ remains sensitive to aspect ratio ($\Gamma$) and flow topology ($\Lambda$) for a much wider range of parameters than $\chi$ (as is the case for other definitions of the Nusselt number, see Appendix \ref{appE}). This result is in agreement with previous studies \citep{Sheard2011} who found no $\Gamma$ dependence using a flux-based definition of the Nusselt number normalized by $\Lambda$ rather than the diffusive solution in the convection-dominated regime{, which spans a large range of $Ra_L$ encompassing our simulation parameters \citep[see also][for the effect of $\Gamma\gg 1$ on the transition from diffusion- to convection-dominated HC]{Hossain2019}}.}

\subsection{Characteristic length scale from auto-correlation}\label{sec:auto}

\begin{figure}
\centering
\includegraphics[width=0.48\textwidth]{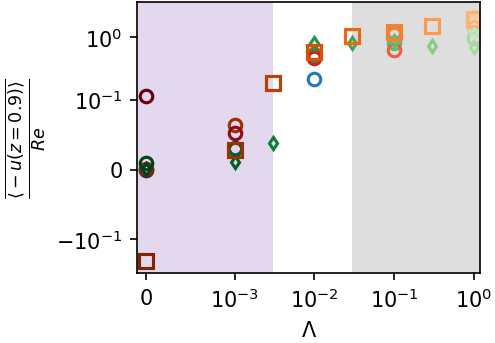}
\includegraphics[width=0.1\textwidth]{grille-de-lecture-des-symboles.png}
\vspace{-0.1in}\caption{Temporally- and horizontally-averaged horizontal flow near the top boundary at $z=0.9$ normalized by the Reynolds number as a function of $\Lambda$. $\overline{\langle -u(z=0.9) \rangle_x}$ is of the same order as $Re$ for relatively large $\Lambda$.}
\label{fig8}
\end{figure}

In this section we estimate the characteristic length scale of the overturning cells in order to further demonstrate that $\Lambda\approx10^{-2}$ marks the transition from RB convection to HC.
Since the leftward flow near the top boundary is an emblematic feature of HC, we use the variations in $x$ of the horizontal velocity at $z=0.9$ as diagnostic.
We first show the time- and $x$-averaged horizontal flow at $z=0.9$ normalized by $Re$ as a function of $\Lambda$ in figure \ref{fig8}. For $\Lambda\geq 10^{-2}$, $\overline{\langle -u(z=0.9) \rangle} \approx Re >0$, which indicates that the large-scale leftward current dominates the dynamics. For small $\Lambda \ll 10^{-2}$, including $\Lambda=0$, $\overline{\langle -u(z=0.9) \rangle} \ll Re$ (either positive or negative) but is not always zero, because a closed domain with moderate aspect ratio can deform  RB-like overturning cells durably, such that a mean flow exists in the upper half of the domain, which is balanced by an equivalently strong return flow in the bottom half. 
Note that the mean horizontal flow exhibits complex fluctuations in time near the transition $\Lambda=10^{-2}$ due to the superposition of the RB and HC dynamics, which cannot be inferred from figure \ref{fig8} but will be discussed in section \S~\ref{sec:hyst}.

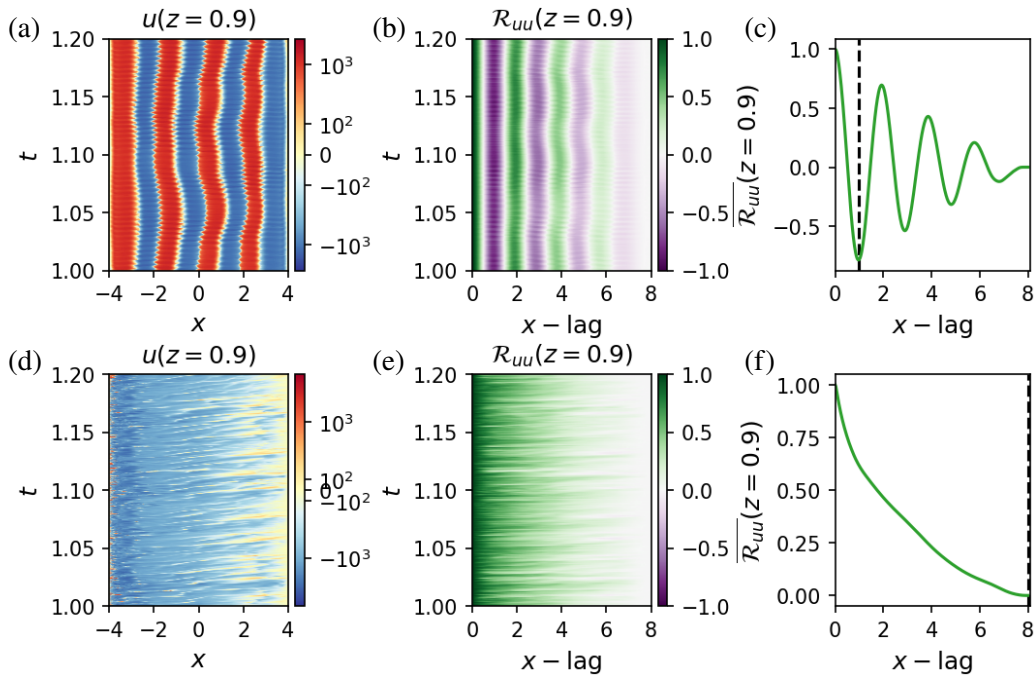
\begin{figure}
\centering
\includegraphics[width=1\textwidth]{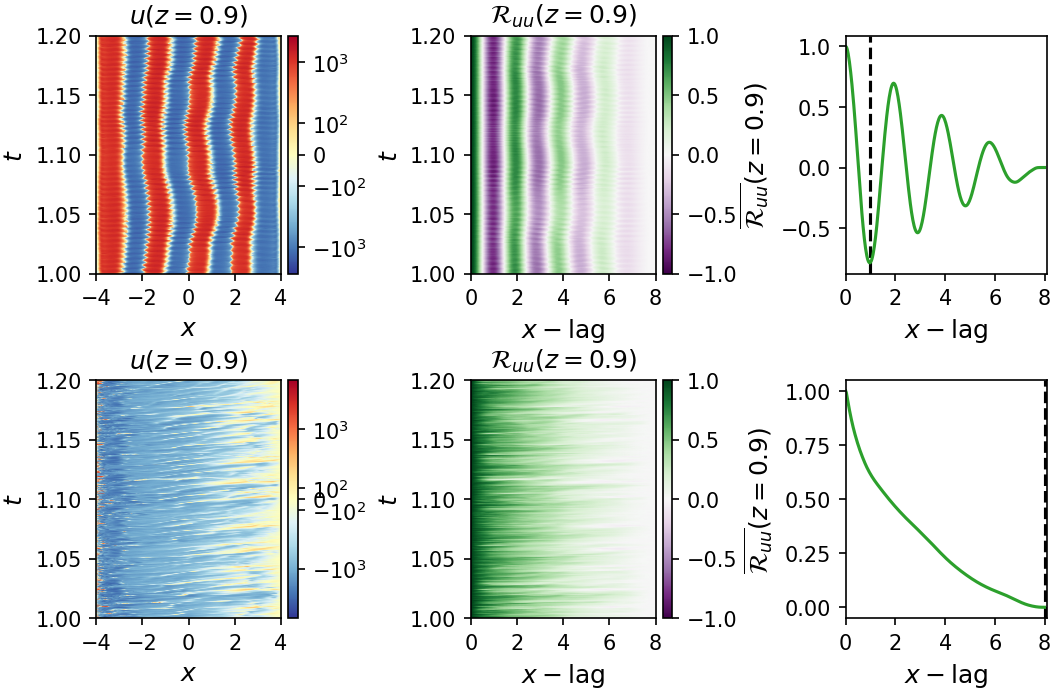}
\put(-387,240){\large{(a)}}
\put(-250,240){\large{(b)}}
\put(-110,240){\large{(c)}}
\put(-387,115){\large{(d)}}
\put(-250,115){\large{(e)}}
\put(-110,115){\large{(f)}}
\vspace{-0.1in}\caption{(a) Horizontal flow near the top boundary ($z=0.9$) as a function of $(x,t)$ for $Ra_F=10^8$, $\Gamma=8$ and $\Lambda=10^{-3}$. (b) Auto-correlation function $\mathcal{R}_{uu}$ in $x$ of the horizontal flow shown in (a) as a function of time $t$. (c) Time average auto-correlation $\overline{\mathcal{R}_{uu}}$ as a function of lag in $x$. {The dashed line shows the first minimum (trough), which we identify as the characteristic length scale of the horizontal flow. (d)-(f) Same as (a)-(c) but for $\Lambda=1$. For a monotonically decreasing auto-correlation function, the characteristic length scale equals the domain length $\Gamma$.}}
\label{fig9}
\end{figure}

The calculation of the characteristic length scale $\ell$ of overturning motions from $u(z=0.9)$ is illustrated in figure \ref{fig9} for a RB-like case ($\Lambda=10^{-3}$; top row) and a HC-like case ($\Lambda=1$; bottom row) with $Ra_F=10^{8}$ and $\Gamma=8$. Figure \ref{fig9}(a) shows the $(x,t)$-Hovmöller diagram of $u(z=0.9)$ for the RB-like simulation at statistical steady state. Four pairs of counter-rotating rolls can be identified, which have approximately the same width and slightly meander in time. The auto-correlation function of $u$ in $x$ is defined as 
\ba{}\label{eq:autocor}
\mathcal{R}_{uu}(x) \equiv \f{\int_{-\Gamma/2}^{\Gamma/2} u(x') u(x'+x) \,\mathrm{d}x'}{\int_{-\Gamma/2}^{\Gamma/2} u(x') u(x')\,\mathrm{d}x'},
\ea
where $u(x'+x>\Gamma/2)$ is set to 0 and $x>0$ is the lag. The auto-correlation function evaluated at $z=0.9$ shows an oscillatory pattern (figure \ref{fig9}(b)), like $u(z=0.9)$, which is damped as the lag increases due to the presence of fluctuations and the scarcity of data for large lag. {We estimate the characteristic length scale $\ell$ of the overturning cells from the first minimum  of the time-averaged auto-correlation function, which is always the largest in absolute value for all our simulations.} {For $\Lambda=10^{-3}$, figure \ref{fig9}(c) shows that $\ell\approx 1$ (dashed solid line), which is approximately the value expected for an unconfined RB roll (rotating clockwise or anti-clockwise)}. Figures \ref{fig9}(d)-(f) show the same results as figure \ref{fig9}(a)-(c) but for $\Lambda=1$. The horizontal flow near the top boundary is now approximately negative everywhere, such that the auto-correlation function monotonically decreases with the lag in $x$. 
 {Thus, the characteristic length scale equals the domain size}, i.e. $\ell=\Gamma$, as shown by the vertical dashed line in figure \ref{fig9}(f). {We note that the auto-correlation function \eqref{eq:autocor} may be defined differently to account for the scarcity of data at large lag (e.g. replacing $u(x')$ with $u(x'+x)$ in the denominator). However, such a definition creates large boundary effects (not shown), which make the calculation of $\ell$ more complicated (because of large variations for large lag), although ultimately unchanged since the locations of most extrema are not modified for small-to-moderate lags.}

\begin{figure}
\centering
\includegraphics[width=0.4\textwidth]{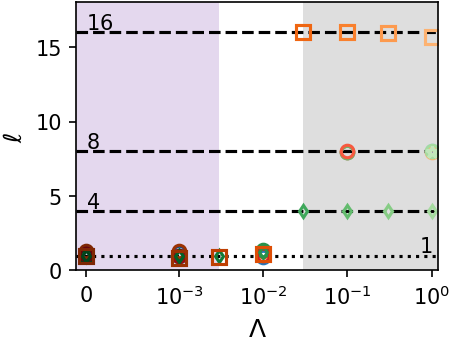}
\includegraphics[width=0.12\textwidth]{grille-de-lecture-des-symboles.png}
\vspace{-0.1in}\caption{Characteristic length scale $\ell$ at statistical steady state as a function of $\Lambda$ for all simulations. The dashed lines show the aspect ratios (or domain lengths), i.e. $\Gamma=4,\; 8$ and $16$. At large $\Lambda$, $\ell=\Gamma$, which is an indication of horizontal convection dominating the dynamics, while at small $\Lambda$, {$\ell\approx 1$ (dotted line)} indicates multiple overturning cells, as in classical RB convection.}
\label{fig10}
\end{figure}

The characteristic length scale $\ell$ obtained from the auto-correlation function of $u(z=0.9)$ is shown for all simulations as a function of $\Lambda$ in figure \ref{fig10}. For $\Lambda \ll 10^{-2}$, $\ell \approx 1$ (dotted line), as expected for classical RB simulations, although there is a small spread between 0.85 and 1.4, which is a result of bounded-box effects or, at non-zero $\Lambda$ values, bursts of large-scale currents sweeping away RB cells (further explained in section \S~\ref{sec:hyst}). For $\Lambda \gg 10^{-2}$, $\ell=\Gamma$, which is here equal to either 4, 8 or 16 (levels shown by dashed lines in figure \ref{fig10}) and an indication that the dynamics is dominated by horizontal convection.

\subsection{Bi-stability and bursts near the transition}\label{sec:hyst}

This section focuses on the transitional regime between the RB and HC dynamical regimes obtained for $\Lambda\ll 10^{-2}$ and $\Lambda\gg 10^{-2}$, respectively.
To this end, we present numerical results that use the method of {discrete} continuation. We fix $Ra_F=10^6$ and $\Gamma=12$, and we gradually vary the flux ratio $\Lambda$ in the range $[0,0.03]$.
For each value of $\Lambda$, we integrate the system for $20$ diffusive timescales to ensure that the system has reached a statistical quasi-steady state.
This integration time is increased up to $100$ diffusive timescales for cases close to bifurcation points, hence limiting this study to only moderately large Rayleigh number and aspect ratio.
In order to track bifurcations between different states, we consider the following averaged horizontal flow
\begin{equation}
    \overline{\langle u\rangle}_{z>0.5}=\frac{2}{\Gamma}\int_{-\Gamma/2}^{\Gamma/2}\int_{0.5}^1 \overline{u} \,\mathrm{d}x \,\mathrm{d}z,
    \label{eq:meanux}
\end{equation}
and mean temperature difference between the right- and left-hand sides of the domain
\begin{equation}
    \Delta T_H=\overline{\langle T\rangle}_{x>\Gamma/4}-\overline{\langle T\rangle}_{x<-\Gamma/4}=\frac{4}{\Gamma}\int_{\Gamma/4}^{\Gamma/2}\int_{0}^1 \overline{T} \,\mathrm{d}x \,\mathrm{d}z-\frac{4}{\Gamma}\int_{-\Gamma/2}^{-\Gamma/4}\int_{0}^1 \overline{T} \,\mathrm{d}x \,\mathrm{d}z,
    \label{eq:meandt}
\end{equation}
where time averaging is only performed once the statistical steady state has been reached.
Due to the temperature profile imposed on the top boundary~\eqref{eq:bcs2}, we expect the time-averaged mean flow~\eqref{eq:meanux} to be generally negative {and the mean temperature difference~\eqref{eq:meandt} to be positive.} Note that by mass conservation, a similar integration as equation \eqref{eq:meanux} performed on the lower half of the domain would yield exactly the opposite mean value.

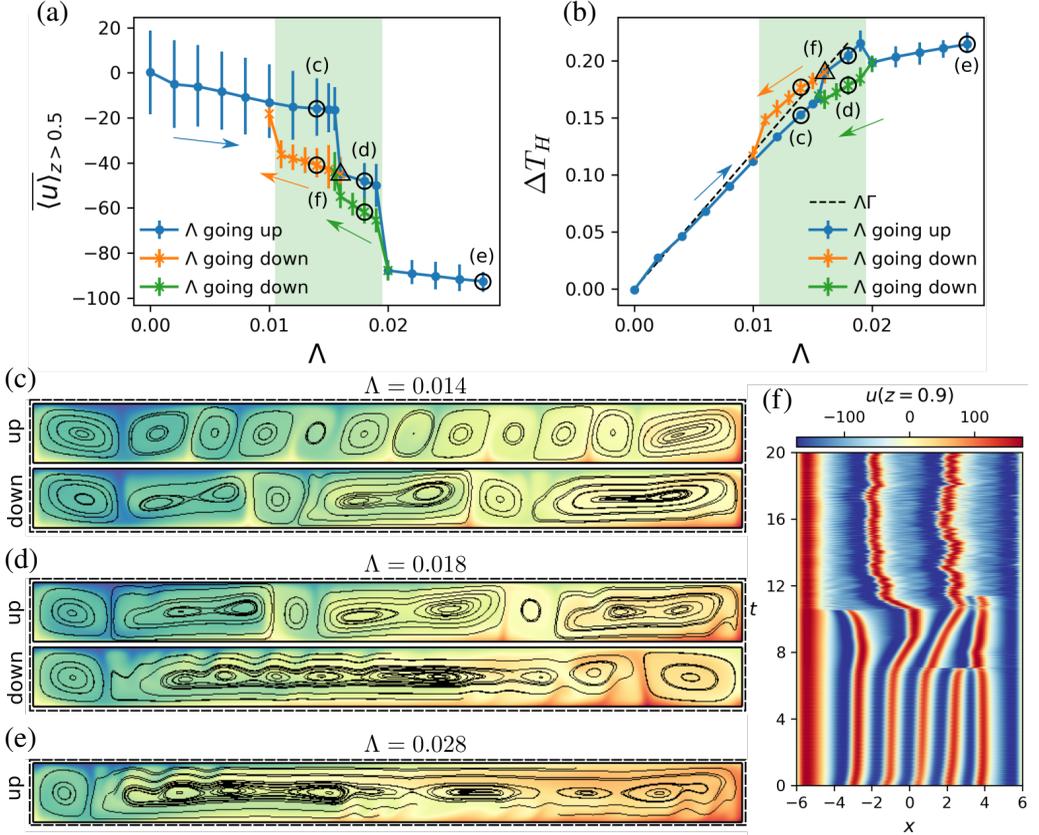
\begin{figure}
\centering
\includegraphics[width=1\textwidth]{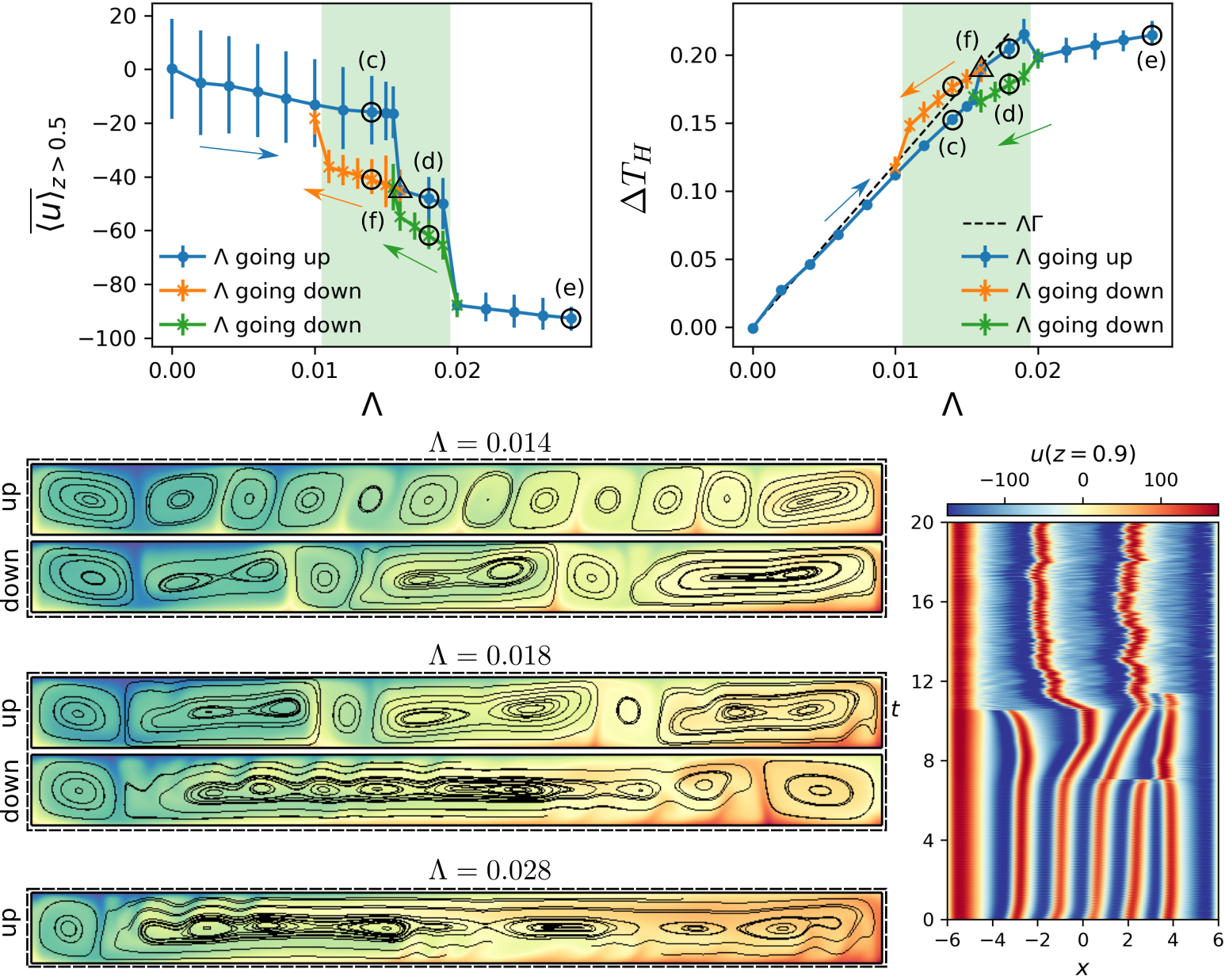}
\put(-373,305){\large{(a)}}
\put(-175,305){\large{(b)}}
\put(-385,168){\large{(c)}}
\put(-385,100){\large{(d)}}
\put(-385,33){\large{(e)}}
\put(-100,160){\large{(f)}}
\vspace{-0.in}\caption{{Bifurcation diagrams based on averaged velocity and temperature.} (a) Mean horizontal flow averaged over time and the upper half of the domain (equation \eqref{eq:meanux}) as a function of the heat flux ratio $\Lambda$. The parameters are $Ra_F=10^6$ and $\Gamma=12$. Round symbols correspond to gradually increasing $\Lambda$ while cross symbols correspond to gradually decreasing $\Lambda$. Each symbol corresponds to a simulation lasting from $20$ and up to $100$ (close to bifurcations) diffusive timescales. The shaded area shows the region of bi-stability. {(b) Same as (a) but for the depth-averaged difference of temperature between the right- ($x>\Gamma/4$) and left-hand  ($x<-\Gamma/4$) sides of the domain (equation \eqref{eq:meandt}).} (c) Snapshots in $(x,z)$ of the temperature field with streamlines shown as black contours for $\Lambda=0.014$ when $\Lambda$ is going up (top panel) and down (bottom panel). (d) Same as (c) but for $\Lambda=0.018$. (e) Same as (c) but for $\Lambda=0.028$ (branch with $\Lambda$ going up only). {(f) Horizontal flow at $z=0.9$ as a function of $x$ and $t$ (initial time set to 0 arbitrarily) for the simulation right after the first transition of the increasing-$\Lambda$ (blue) branch on panels \ref{fig11}(a),(b) ($\Lambda=0.016$; highlighted by triangles). Note that merging events, e.g. at $x\approx 2$ and $t \approx 7$, appear discontinuous because they occur on short time scales.}}
\label{fig11}
\end{figure}

We start by gradually increasing $\Lambda$ from 0 to $0.03$.
The results are shown in figure~\ref{fig11}(a),(b) as round symbols.
We recall that each point corresponds to at least $20$ diffusive timescales and up to $100$ timescales for cases on each side of a given transition.
The mean horizontal flow in the upper half of the fluid domain gradually increases in absolute value until a first transition occurs at $\Lambda\approx0.016$ (figure~\ref{fig11}(a)).
This transition corresponds to the first destabilisation of the Rayleigh-B\'enard cells with approximately unit aspect ratio, which merge into horizontally elongated cells.
For our particular choice of $\Gamma=12$, the flow is initially organised in twelve cells and switches to six more elongated cells.
A secondary transition between these six cells and only one extended cell is observed at $\Lambda=0.02$.
The system remains in this state for larger values of $\Lambda$.
{The two transitions {at $\Lambda=0.016$ and $\Lambda=0.02$} are also observed in the mean horizontal temperature difference (figure \ref{fig11}(b)). 
The mean horizontal temperature difference $\Delta T_H$ initially follows the temperature difference imposed along the top boundary, which is equal to $\Lambda\Gamma$ (dashed line in figure~\ref{fig11}(b)). As $\Lambda$ increases, $\Delta T_H$ slowly deviates downward until it recovers the imposed value $\Lambda\Gamma$ just after the transition at $\Lambda=0.016$.
After the second transition, $\Delta T_H$ clearly follows a different trend {and increases less rapidly with $\Lambda$}, which is indicative of the growing efficiency of HC in its ability to mix the imposed horizontal temperature difference.
}

To explore the possibility of multi-stability, we gradually decrease $\Lambda$ from a given equilibrium state.
We do so {independently} for each observed transition, i.e. {starting first} at $\Lambda=0.016$ and then at $\Lambda=0.02$.
We show the corresponding descending branches in figure~\ref{fig11}(a),(b) using cross symbols.
We clearly observe hysteretic behaviour over a large range of heat flux ratio $\Lambda$, which is highlighted by the green shaded region in figure~\ref{fig11}(a),(b).
The bi-stability is characterized by the separation of the ascending and descending branches, which are each connected to distinct flow organizations: for $0.01<\Lambda<0.02$, there always exist at least two flow states for the same $\Lambda$ that have a different number of convective rolls; each roll having an aspect ratio that can vary between 1 and $\Gamma$. Thus, the actual number of convective cells in both horizontal and vertical directions crucially depends on the history of the system. The possibility to have different number of rolls for the same $\Lambda$ can be seen from the lower-left panels on figure~\ref{fig11} where we show the temperature field and streamlines for two bi-stable states at $\Lambda=0.014$ (figure~\ref{fig11}(c)) and $\Lambda=0.018$ (figure~\ref{fig11}(d)). For completeness, we also show in figure~\ref{fig11}(e) an example of the final state of pure horizontal convection reached once $\Lambda>0.02$.
{
While the mean flow is always stronger for descending than for ascending branches, the number of rolls has a non-trivial effect on the mean horizontal temperature difference, since $\Delta T_H$ can either increase (orange line above blue line in figure~\ref{fig11}(b)) or decrease (blue line above green line in figure~\ref{fig11}(b)) with decreasing roll number. The former behavior may be explained from the persistence of HC dynamics along the descending branch (green line), which efficiently mixes the horizontal temperature gradient, whereas the latter suggests, counter-intuitively, that, for our choice of parameters, six rolls are more efficient at maintaining a large $\Delta T_H$ than twelve rolls.
}

{We would like to make a few notes regarding the results of figures \ref{fig11}(a)-(e). First, the merging of rolls does not percolate from either the left- or right-hand side of the domain, where HC drives distinct dynamics (intense downwelling versus weak upwelling). Rather, it occurs in the bulk through nucleation, as shown e.g. in figure~\ref{fig11}(f) wherein the first merging occurs between the second and third pair of rolls from the right boundary. Second, for low values of $\Lambda<0.01$, we still observe bi-stability since the system does not recover exactly its initial state (characterizing the increasing branch) as we continue decreasing $\Lambda$. However, the two solutions only differ by the spatial organisation of the convective rolls, not their numbers. Third, we do not observe tri-stability close to $\Lambda\approx0.016$: the ascending trajectory bifurcates exactly where the trajectory descending from $\Lambda=0.02$ falls back onto the descending trajectory initialised from $\Lambda=0.016$. This does not mean that tri-stability cannot be obtained in RBH convection systems. In fact, we expect the bifurcation diagram to be richer than what can be deciphered by our study, especially as $\Gamma$ increases. Finally, even though we have integrated these bi-stable states for up to $100$ diffusive timescales based on the height of the fluid domain, we cannot exclude the possibility that rare spontaneous transitions can occur between them (partly because the diffusive timescale based on the horizontal size of the domain scales with $\Gamma^2\gg 1$, hence is much longer). One way to more firmly prove that bi-stable states exist would require a detailed measurement of the transition time between different states, as was recently done in thin-layer turbulence \citep{dewit_2022}, which is beyond the scope of the present study.}

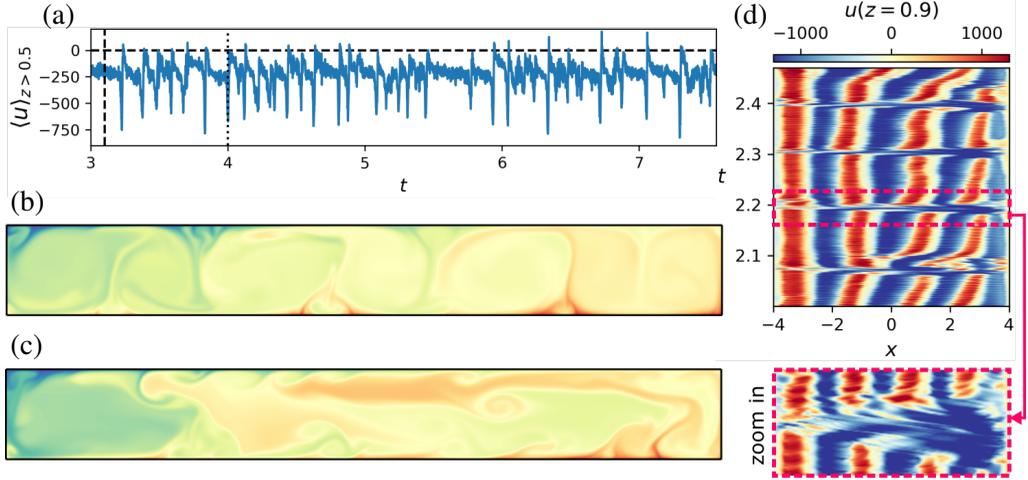
\begin{figure}
\centering
\includegraphics[width=1\textwidth]{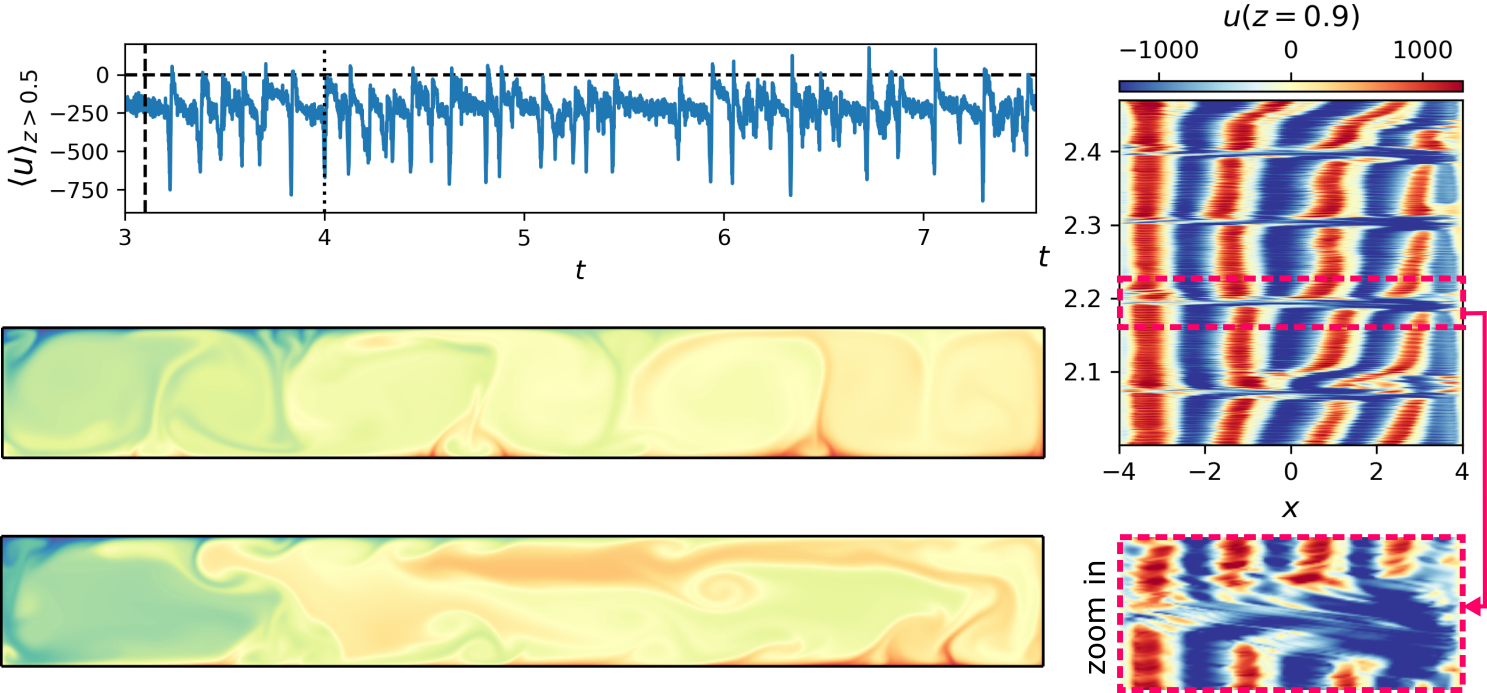}
\put(-370,170){\large{(a)}}
\put(-382,100){\large{(b)}}
\put(-382,46){\large{(c)}}
\put(-110,170){\large{(d)}}
\vspace{-0.in}\caption{(a) Mean horizontal flow averaged over the upper half of the domain as a function of time for $Ra_F=10^8$, $\Gamma=8$ and $\Lambda=10^{-2}$. (b), (c) Snapshots of the temperature field at times $t=3.1$ (vertical dashed line in the top panel) and $t=4$ (vertical dotted line), respectively. {(d) Spatio-temporal plot of $u(z=0.9)$ for the same simulation.}}
\label{fig12}
\end{figure}

We expect the hysteretic behaviour to persist for aspect ratios larger than $\Gamma=O(10)$, since this allows for even more distinct convective cell configurations (see \cite{Wang2020} for classical RB convection).
However, it is  unclear whether such dynamics will persist at large Rayleigh numbers.
The analysis of bi-stable dynamics at large $Ra_F$ is obviously very intensive numerically as it requires integrating a fully turbulent system for hundreds of diffusive timescales.
Nevertheless, we explore a particular case close to a transition to show that, while we do not observe bi-stability, we do observe bi-modality in certain range of parameters.
For example, let us consider the much more turbulent case $Ra_F=10^8$ and $\Gamma=8$.
For $\Lambda=0.01$, i.e. right at the lower bound of the bi-stable regime observed at $Ra_F=10^6$, we observe spontaneous transitions between states with large and weak horizontal mean flows, which can be seen from the time history of the horizontal mean flow (defined in equation \eqref{eq:meanux}) in figure~\ref{fig12}(a). The bi-modality is clear: the horizontal mean flow displays abrupt (negative) HC-like peaks before rapidly relaxing to a state of weaker RB-like mean flow.
{The mean-flow bursts correspond to abrupt intrusions of warm fluid accumulating below the heated part of the right corner, which can temporarily disrupt the Rayleigh-B\'enard cells, and are reminiscent of the burst dynamics observed in RB convection between stress-free plates \citep{Goluskin2014}.}
The two bottom panels of figure~\ref{fig12} show snapshots of the temperature field of a single simulations exhibiting bi-modality: figure \ref{fig12}(b) displays convective rolls reminiscent of classical RB convection, while figure \ref{fig12}(c) shows a state featuring a burst of horizontal convection (note that a similar HC burst can be seen in the movie available as Supplementary Material). {Figure \ref{fig12}(d) shows the time history of $u(z=0.9)$, which can help visualize the horizontal development of the bursts: the bursts always start from the right boundary, then the system relaxes from the left boundary.} 
For this particular value of $\Lambda$ and duration of the simulation, the convection cells always recover their RB-like configuration and HC remains intermittent.
Whether the system remains in this bi-modal state or eventually converge towards one of the two attractors at long times is an open question. 

\begin{figure}
\centering
\includegraphics[width=0.95\textwidth]{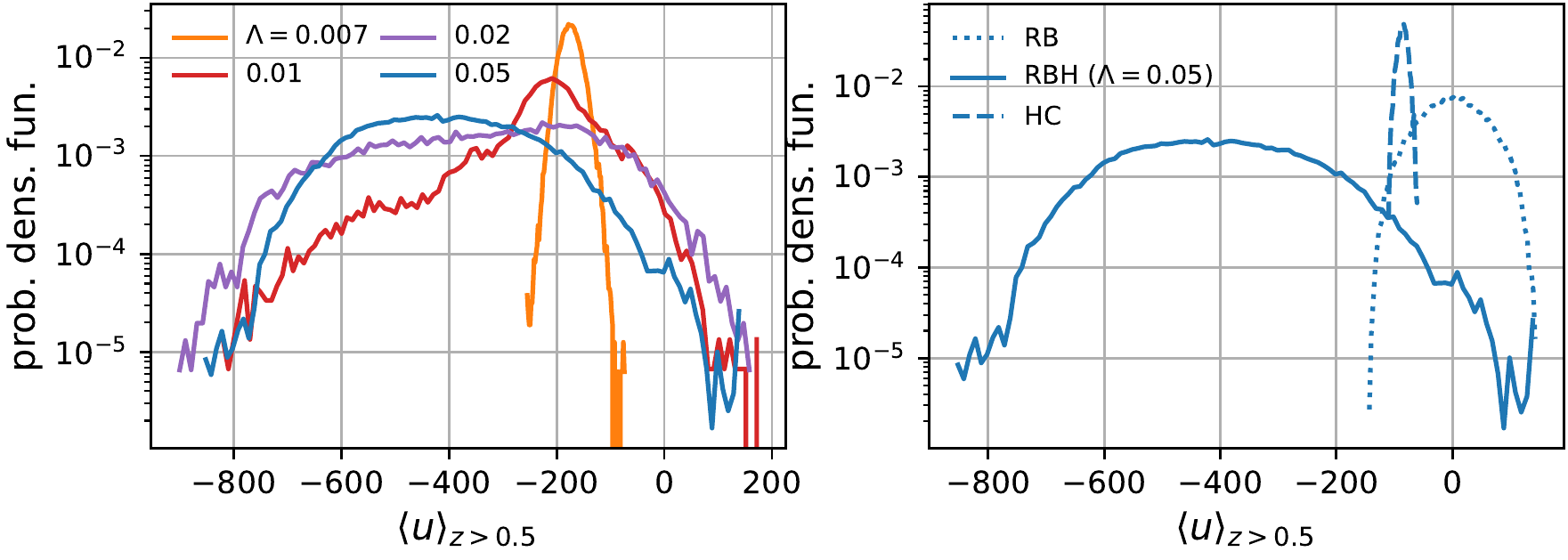}
\put(-350,128){\large{(a)}}
\put(-170,128){\large{(b)}}
\vspace{-0.1in}\caption{(a) Probability density function (PDF) at statistical steady state of {${\langle u \rangle}_{z>0.5}$} for $Ra_F=10^8$, $\Gamma=8$ with $\Lambda=0.007$, 0.01, 0.02, 0.05. (b) Same as (a) but for pure RB convection ($\Lambda=0$; dotted line), pure HC ($\Lambda=0.05$ but no geothermal flux; dashed line), and RBH convection ($\Lambda=0.05$; solid line).}
\label{fig13}
\end{figure}

We conclude our analysis of RBH dynamics by discussing and comparing the Probability Density Function (PDF) of the horizontal mean flow defined by equation~\eqref{eq:meanux} in different regimes (including bi-modality) near the transition. We consider fixed (relatively large) $Ra_F=10^8$, $\Gamma=8$ and variable $\Lambda$ close to the transition, as well as pure RB convection ($\Lambda=0$) and pure HC (no geothermal flux) for comparison.
We first show in figure~\ref{fig13}(a) the mean flow PDF at statistical steady state for simulations close to the transition.
We clearly see an abrupt transition between a Gaussian distribution for $\Lambda=7\times10^{-3}$ and a skewed distribution with an elongated negative tail for $\Lambda=10^{-2}$.
As $\Lambda$ further increases, the PDFs become more and more skewed toward large negative values, but maintain a large spread confirming the intermittent nature of the mean flow dynamics.
Interestingly, the burst dynamics is a specific property of the mixed RB-HC regime, i.e. which is absent in pure RB or HC regimes.
Figure~\ref{fig13}(b) shows the PDF of the mean horizontal upper flow for three cases at $Ra_F=10^8$ and $\Gamma=8$: one with $\Lambda=0$ (pure RB case), one with $\Lambda=0.05$ but no geothermal flux (pure HC) and finally the RBH case with $\Lambda=0.05$ and geothermal heating.
We clearly see that both RB and HC cases produce mean-flow PDFs that are approximately Gaussian and centered around zero (dotted line) and a small negative value (dashed line), respectively.
Conversely, the RBH case yields a PDF with a large spread, which is consistent with rare but intense mean-flow bursts and thus a characteristic feature of RBH dynamics.
The emergence of strong mean flows (persistent or bursty) in RBH convection is consistent with earlier results \citep{Mullarney2006} and confirms the underlying tendency of RB convection to drive large-scale horizontal flows, which are here enhanced by the imposed horizontal temperature gradient.
One might expect that the bursts eventually disappear at large $\Lambda$, i.e. once the mean flow driven by the upper horizontal temperature gradient becomes dominant compared to the maximum value observed during each burst event. The underlying $\Lambda$ threshold would correspond to the lower bound of the HC-dominated regime, highlighted by grey shadings in figures \ref{fig7a}, \ref{fig7b}, \ref{fig8} and \ref{fig10}. We leave the detailed investigation of this threshold, which will require exploration of the burst dynamics in the large $Ra_F$ and large $\Gamma$ limit, to future studies.

\section{Discussion and conclusions}\label{sec:conc}

We have investigated the dual RBH convection problem via direct numerical simulations in order to identify the transition from RB convection to HC in parameter space. The $\Lambda$ parameter, which is the ratio of the top horizontal heat flux divided by the bottom heat flux, clearly controls the system dynamics: RBH transitions from RB convection at small $\Lambda\leq 3\times 10^{-3}$ (purple shading in figures \ref{fig7a}, \ref{fig7b}, \ref{fig8} and \ref{fig10}) to HC at large $\Lambda\geq 3\times 10^{-2}$ (grey shading). Importantly, the transition occurs near $\Lambda=10^{-2}$ independently of the aspect ratio $\Gamma$ and Rayleigh number $Ra_F$, which serves as a proxy for energy input from both the bottom and top boundary.

{The Nusselt numbers of RB convection and HC (defined in equations \eqref{eq:nurb} and \eqref{eq:nuhc}) are good indicators of the flow regime; they deviate quickly from the classical scaling laws of RB convection and HC in the regime where they are not relevant, i.e. $\Lambda\gg 10^{-2}$ for $Nu_{RB}$ and $\Lambda\ll 10^{-2}$ for $Nu^{\chi}_{HC}$. The characteristic length scale $\ell$ derived from the auto-correlation function (\S~\ref{sec:auto}) is another excellent indicator as it is equal to 1 in the RB regime and $\Gamma$ in the HC regime. The Reynolds number is not a good indicator for our set of simulations, however, because it is primarily controlled by $Ra_{F}$, the intensity of buoyancy driving, with small variations only due to $\Lambda$ (flow topology) and $\Gamma$.}

{The fact that the transition does not depend on $\Gamma$ (at least for $\Gamma\geq 4$, as considered in this study) means that the competition mechanism is mostly local and does not depend on the horizontal extent nor the number of convective rolls present.
Since we observe intense mean flows near the transition, i.e. more intense than those observed in pure RB or HC regimes at similar parameter values (the results of figure~\ref{fig13}(b) are for $Ra_F=10^8$ but are qualitatively representative of simulations using other $Ra_F$), we suspect that the Reynolds stresses originating from the convective rolls are enhanced by the imposed horizontal temperature gradient (by favouring counter-clockwise rolls), leading to efficient mean flow amplification.
Research on horizontal mean flows coupling with RB convection cells in periodic domains has a long history \citep{Thompson1970,Krishnamurti1981,Busse1983}.
For instance, arrays of convective rolls are known to be unstable to the so-called shearing instability \citep{Hughes1990,Rucklidge1996,Goluskin2014} when both horizontal plates are stress free, whereby Reynolds stresses positively couple with the roll tilt induced by the emerging mean flow. We do not expect to directly observe this shear instability, especially for $\Lambda=0$, because of our closed domain and no-slip boundary conditions, which are known to inhibit the instability for elongated domains \citep{Fitzgerald2014,VanDerPoel2014}. However, for $\Lambda>0$, we suspect that the imposed symmetry breaking in the horizontal direction activates a mechanism reminiscent of the shearing instability, which might explain why a weak horizontal temperature gradient can easily disrupt the convective rolls.
The subtle interplay between the horizontal mean flow triggered by the imposed temperature gradient along the top boundary and the convective rolls  most likely underpins the origin of the transition from RB convection to mixed RBH convection, as well as the complex bi-stable and bi-modal dynamics observed at $\Lambda\approx 10^{-2}$. Therefore, it would be of interest to revisit the secondary instability of convection rolls in the presence of an imposed horizontal temperature gradient driving a mean flow using e.g. a weakly-nonlinear theory in future work.}

We have shown that the system is multi-stable near the transition. For $Ra_F=10^6$, we have found at least two stable branches for $0.01\leq\Lambda\leq 0.02$. Each branch corresponds to a different number of overturning cells. Interestingly, we  found that multiple flow states can also exist for a fixed number of rolls, as the spatial organization of the rolls can differ substantially between cases. For larger $Ra_F=10^8$, the bi-stability seems to disappear, at least for simulations lasting only a few diffusive time scales, possibly because turbulent fluctuations force transitions between the two states so efficiently that their basin of attraction overlap. For $Ra_F=10^8$ and $\Lambda=10^{-2}$ the dynamics is better described as bi-modal, i.e. akin to RB convection with bursts of HC. 

Obviously, the independence of the transition observed at $\Lambda=10^{-2}$ with $Ra_F$ is rigorously valid only for $10^6 \leq Ra_F\leq 10^9$ as considered in this study. Exploring the RBH dynamics at larger $Ra_F$ and in three dimensions is thus required to extend the applicability of our results to geophysical fluids, such as Earth's atmosphere and subglacial lakes. The Prandtl number for the latter is $Pr=O(10)$, hence is much larger than $Pr=1$, but its effect on the transition may be limited. Indeed, increasing $Pr$ for $Pr \geq 1$ has little effect on $Nu_{RB}$ and $Nu^{\chi}_{HC}$, whereas $Re$ decreases almost like $Pr^{-1}$ in both RB convection and HC, at least for moderate Rayleigh numbers \citep{Shishkina2016b,Li2021}. Thus, we hypothesize that increasing $Pr$ may slow down fluid motions without changing the type of convection. That being said, we note that the burst dynamics in RB convection between stress-free plates has been observed to disappear for large enough $Pr$ \citep{Goluskin2014}, suggesting that the upper bound of the RB regime (rightmost edge of purple shading in, i.e. figure \ref{fig10}) may be in fact sensitive to $Pr$.

The dual {Rayleigh--Bénard}-Horizontal convection problem is likely to receive increased attention in the coming years because subglacial lakes will soon be explored and monitored (including, notably, lake CECS; see \citet{Rivera2015}) and because the contribution of geothermal heating to the ocean abyssal stratification and circulation is now known to be significant \citep{Mashayek2013,DeLavergne2016}. For subglacial lakes, our results suggest that RB convection should dominate since $\Lambda\ll 10^{-2}$ for realistic ice-water interface slopes (Appendix \ref{appB}). {It is unclear whether Rayleigh--Bénard convection will transition toward HC at lower or higher $\Lambda$ in three dimensions than in two dimensions. Preliminary three-dimensional simulations suggest a transition around $\Lambda\approx 10^{-2}$ again, which means that our results may be (at least qualitatively) informative for real systems in spite of the two-dimensional limitation.} Non-rectangular geometry and Earth's rotation, whose impact on subglacial lake dynamics remains unclear \citep{Couston2021b}, are other effects not considered in this work that may be more favorable to HC, hence should be investigated in future studies. The nonlinear equation of state of freshwater is also an important physical ingredient that can completely modify the fluid dynamics of subglacial lakes \citep{Couston2021,Olsthoorn2021}, which would be interesting to consider in RBH convection.

{Despite the numerous approximations of our work, we provide in closing some thoughts on the implications of the fluid dynamics regime (considering both RB convection and HC) for the future exploration of subglacial lakes. 
For simplicity, our discussion assumes flat water-bedrock interface, constant and positive thermal expansion coefficient and neglects the effect of rotation. In addition to measuring flow velocities, temperature and salinity along vertical profiles, future explorations of subglacial lakes will most likely investigate populations of suspended particulates (including microorganisms) through direct sampling of the water column and analysis of accreted ice (which may host particulates due to freezing of the lake water onto the ice ceiling), and characterize past climates by analyzing sediment cores extracted from the lake bed \citep{Hodgson2009}. In lakes dominated by HC, vertical profiles should be ideally performed at both ends of the lake in order to probe both the downwelling and upwelling regions. The former may be affected by subglacial water discharge from upstream (if any) whereas the latter is best to search for suspended particulates, whether in the water column or in the ice above. Sediments brought in the lake via upstream subglacial discharge will likely accumulate below the downwelling region. Dating sediments from continuous cores (resulting from successive layer depositions) is typically easier than from non-continuous cores \citep{Hodgson2009}. Thus, initial dating could be based on cores extracted from below the downwelling region, henceforth enabling dating sediment cores extracted from below the upwelling region that are more ancient. In lakes dominated by RB convection, the lack of upstream/downstream asymmetry combined with the possible migration of RB cells means that the drilling strategy can be based on constraints other than the water circulation, which homogenizes lake conditions. In this case, four drilling sites separated (horizontally) by a quarter water depth along the main direction of the lake would most likely provide good coverage of upwelling and downwelling regions.}


\backsection[Acknowledgements]{We gratefully acknowledge support from the PSMN (Pôle Scientifique de Modélisation Numérique) of the ENS de Lyon for the computing resources. Centre de Calcul Intensif d'Aix-Marseille is acknowledged for granting access to its high-performance computing resources. This work was performed using HPC/AI resources from GENCI-IDRIS/TGCC (Grant 2021-A0120407543). The authors would like to thank Adrien Villaret for his careful review and Andy Smith and Keith Makinson for useful discussions about subglacial lake exploration.}

\backsection[Funding]{This research received funding from LabEx LIO (ANR-10-LABX-0066), Université de Lyon.}

\backsection[Declaration of interests]{The authors report no conflict of interest.}




\appendix

\section{Simulations details}\label{appA}

The physical and numerical parameters of all simulations that led to the results discussed in sections \S~\ref{sec:phen}-\ref{sec:auto} are provided in table \ref{tab:sims}. The details of the simulations used for the multi-stability analysis are provided in \S~\ref{sec:hyst} and summarized on the last line of table \ref{tab:sims}.

\begin{table}\centering 
\begin{tabular}{C{1cm}C{1.5cm}C{1cm}C{1.4cm}C{1.6cm}C{1cm}C{1cm}C{1.2cm}C{1cm}}
\cmidrule{1-9}
$\Gamma$ & Geo. Flux & $Ra_F$ & $\Lambda$ & $Ra_L$ & $n_z$ & $l_d$ & $dt \times 10^6$ & Note  \\ 
8 & yes & $10^6$ & $0$ & $0$ & 10 & 7 & 9 & DNS \\
- & - & - & $10^{-3}$ & $\approx 4 \times 10^6$ & - & - & 9 & - \\
- & - & - & $10^{-2}$ & $\approx 4 \times 10^7$ & - & - & 7 & - \\
- & - & - & $10^{-1}$ & $\approx 4 \times 10^8$ & - & - & 5 & - \\
- & - & - & $1$ & $\approx 4 \times 10^9$ & - & - & 2 & - \\
- & no & - & $1$ & $\approx 4 \times 10^9$ & - & - & 2 & - \\
- & yes & $10^7$ & $0$ & $0$ & - & 9 & 2 & - \\
- & - & - & $10^{-3}$ & $\approx 4 \times 10^7$ & - & - & 2 & - \\
- & - & - & $10^{-2}$ & $\approx 4 \times 10^8$ & - & - & 1 & - \\
- & - & - & $10^{-1}$ & $\approx 4 \times 10^9$ & - & - & 1 & - \\
- & - & - & $1$ & $\approx 4 \times 10^{10}$ & - & - & 0.5 & - \\
- & no & - & $1$ & $\approx 4 \times 10^{10}$ & - & - & 0.5 & - \\
- & yes & $10^8$ & $0$ & $0$ & - & 11 & 0.4 & - \\
- & - & - & $10^{-3}$ & $\approx 4 \times 10^8$ & - & - & 0.4 & - \\
- & - & - & $10^{-2}$ & $\approx 4 \times 10^9$ & - & - & 0.3 & - \\
- & - & - & $10^{-1}$ & $\approx 4 \times 10^{10}$ & - & - & 0.2 & - \\
- & - & - & $1$ & $\approx 4 \times 10^{11}$ & 20 & 7 & 0.1 & - \\
- & no & - & $1$ & $\approx 4 \times 10^{11}$ & - & 11 & 0.04 & LES \\
- & yes & $10^9$ & $0$ & $0$ & - & 7 & 0.2 & DNS \\
- & - & - & $10^{-3}$ & $\approx 4 \times 10^9$ & - & - & 0.2 & - \\
- & - & - & $10^{-2}$ & $\approx 4 \times 10^{10}$ & - & - & 0.1 & - \\
- & - & - & $10^{-1}$ & $\approx 4 \times 10^{11}$ & 40 & 11 & 0.02 & LES \\
4 & - & $10^8$ & $0$ & $0$ & 20 & 7 & 0.5 & DNS \\
- & - & - & $10^{-3}$ & $\approx 2.5 \times 10^7$ & - & - & 0.5 & - \\
- & - & - & $3\times 10^{-3}$ & $\approx 7.5 \times 10^7$ & - & - & 0.5 & - \\
- & - & - & $10^{-2}$ & $\approx 2.5 \times 10^8$ & - & - & 0.5 & - \\
- & - & - & $3\times 10^{-2}$ & $\approx 7.5 \times 10^8$ & - & - & 0.4 & - \\
- & - & - & $10^{-1}$ & $\approx 2.5 \times 10^9$ & - & - & 0.5 & - \\
- & - & - & $3\times 10^{-1}$ & $\approx 7.5 \times 10^9$ & - & - & 0.1 & - \\
- & - & - & $1$ & $\approx 2.5 \times 10^{10}$ & 40 & - & 0.1 & - \\
16 & - & $10^7$ & $0$ & $0$ & 20 & - & 1 & - \\
- & - & - & $10^{-3}$ & $\approx 6.5 \times 10^8$ & - & - & 1 & - \\
- & - & - & $3\times 10^{-3}$ & $\approx 2 \times 10^9$ & - & - & 1 & - \\
- & - & - & $10^{-2}$ & $\approx 6.5 \times 10^9$ & - & - & 1 & - \\
- & - & - & $3\times 10^{-2}$ & $\approx 2 \times 10^{10}$ & - & - & 0.6 & - \\
- & - & - & $10^{-1}$ & $\approx 6.5 \times 10^{10}$ & - & - & 0.5 & - \\
- & - & - & $3\times 10^{-1}$ & $\approx 2 \times 10^{11}$ & - & - & 0.4 & - \\
- & - & - & $1$ & $\approx 6.5 \times 10^{11}$ & - & - & 0.1 & - \\
12 & yes & $10^6$ & $0\leftrightarrow 0.03$ & $0\leftrightarrow 6\times10^8$ & $10$ & $7$ & $7$ & DNS \\
\hline
\end{tabular}\vspace{-0.in}\caption{Physical and numerical parameters of the simulations, with a dash denoting same value as in the row above for readability. $\Gamma$ is the aspect ratio, $Ra_F$ is the flux-based Rayleigh number, $\Lambda$ is the heat flux ratio, $Ra_L$ is the horizontal Rayleigh number, $n_z$ is the number of elements in the vertical direction, $l_d$ is the polynomial order and $dt$ is the typical time step at statistical steady state. Most simulations include a bottom heat flux, also referred to as geothermal heating (Geo. Flux), and are run via direct numerical simulation (DNS) from start to finish; some of the most demanding simulations are run via Large-Eddy Simulation (LES) using the filtering approach described in \citet{Fischer2001} from $t=0$ to $t=1$ and then restarted with numerical parameters as indicated in the table and integrated from $t=1$ to $t=1.2$ via DNS. The last line shows the parameters of the simulations discussed in section~\ref{sec:hyst}, which focus on bi-stability.}\label{tab:sims}\end{table}

\section{Parameter range relevant to subglacial lakes}\label{appB}

In this Appendix we motivate the choice of small $\Lambda$ considered in the paper. We are particularly interested in subglacial lakes, which lie beneath several kilometers of ice in Antarctica and Greenland and are subject to geothermal heating and horizontal temperature gradients along their bottom and top boundaries, respectively \citep{Livingstone2022}. The latter arises when variable ice thickness above the lake water results in a tilted ice-water interface by hydrostatic equilibrium. The temperature of freezing varies along a tilted ice-water interface because it depends on local pressure, which is larger where the ice is thicker \citep{Thoma2010a}. A linear approximation for the freezing temperature of freshwater with no dissolved air as a function of local ice pressure $p_i$ (in decibar) or ice thickness $h_i=10^4p_i/(\rho_ig)$ (in meter), with $\rho_i=917$ kg/m$^3$ the ice density and $g=9.81$ m/s$^2$ the surface gravity, is 
\ba{}\label{appBeq1}
T_f = 6.67\times 10^{-2}-8.21\times 10^{-4} p_i = 6.67\times 10^{-2}-9.12\times 10^{-4} h_i, 
\ea
as obtained from best-fit with the exact freezing temperature calculated with the python package gsw \citep{McDougall2011} for $0<p_i<5000$ dbar ($h_i$ ranging from 0 to 5558 m). Equation \eqref{appBeq1} shows that $T_f$ decreases with $p_i$ or $h_i$ ($T_f\approx -4 \; ^{\circ}$C when $p_i=5000$ dbar) and yields a horizontal temperature gradient equal to
\ba{}
\lambda = \f{dT_f}{dx} = 9.12\times 10^{-4} \gamma, 
\ea
for a subglacial lake with an ice-water interface slope $dh_i/dx = -\gamma $, i.e. thicker for smaller $x$, as assumed in this study. We find that $\gamma=0.016$, 0.003, 0.03, 0.002, 0.003, for the five well-documented subglacial lakes reported in \citet{Couston2021b}, i.e. CECs, SPL, Ellsworth, Vostok and Concordia, respectively. This range of $\gamma$ values yields approximately $3\times 10^{-5}<\Lambda=k\lambda/F<3\times 10^{-4}$, which lies within the range $0\leq \Lambda \leq 1$ explored in this study, with $k=0.56$ W/(m $^{\circ}$C) the thermal conductivity of water and $F=50$ mW/m$^2$ an average value for Earth's geothermal flux. Thus, for subglacial lakes, $\Lambda\ll 10^{-2}$, unless the slope is of order unity (in which case $\Lambda\geq 10^{-2}$), which is most likely unrealistic. For completeness, the flux-based Rayleigh number of a subglacial lake below thick ice with depth $H=100$ m, $F=50$ mW/m$^2$, $\nu=1.7\times 10^{-6}$ m$^2$/s, $\kappa=1.33\times 10^{-7}$ m$^2$/s, $\alpha=10^{-4}\; ^{\circ}$C$^{-1}$ (typical values inferred from \citet{Couston2021}) is $Ra_F\approx 3.9 \times 10^{16}$, which cannot be reached in numerical simulations, even in two dimensions, with present-day computational resources.

\section{Scaling laws in the RB and HC regimes}\label{appC}

In this section we comment on the scaling laws inferred from our simulations in light of previously published results (see table \ref{tab:fit}).

The $Re(Ra_F)$ scaling can be recast as a $Re(Ra_{\Delta})$ scaling, where $Ra_{\Delta}$ is the classical {Rayleigh--Bénard parameter based on the temperature difference $\Delta$ between the top and bottom plates (averaged in $x$ and $t$ and computed \textit{a posteriori})}. To see this, we first remark that the different definitions of the problem parameters yield the relationship (in the limit $\Lambda\rightarrow 0$) $Nu_{RB}=Ra_F/Ra_{\Delta}$ \citep{Johnston2009}, such that $Ra_F=a_{RB}^{1/(1-b_{RB})}Ra_{\Delta}^{b_{RB}/(1-b_{RB})}$. This yields the scaling law $Nu=0.316Ra_{\Delta}^{0.236}$ using the pre-factor and exponent reported in table \ref{tab:fit}. {The obtained $Nu(Ra_{\Delta})$ scaling law is in relatively good agreement with \citet{Grossmann2000}'s unifying theory: the exponent is slightly larger (and the pre-factor is slightly smaller) than that predicted by regime $II_u$, i.e. 1/5, which is expected for our low $1.8\times 10^5\leq Ra_{\Delta}\leq 4.8\times 10^7$ numbers \citep[see updated regime diagram in][and note that our scaling law happens to be in better agreement with the 1/4 exponent of regime $I_u$ expected at larger $Pr>1$]{Stevens2013}. We expect the difference between our scaling exponent and that of regime $II_u$ to be due to the two-dimensional flow assumption, which is known to affect $Nu_{RB}$, especially at low-to-moderate $Pr$ \citep{VanDerPoel2013} and the relative proximity of regime $IV_u$ at slightly larger $Ra_{\Delta}$, which features a 1/3 exponent. We note that our exponent is smaller (and the pre-factor is larger) than other two-dimensional flux-based RB studies \citep{Johnston2009,Couston2021} (compare 0.236 with $2/7\approx 0.286$) because we considered lower Rayleigh numbers.} Replacing $Ra_F$ with $Ra_{\Delta}$ in $Re(Ra_F)$ yields $Re=0.140Ra_{\Delta}^{0.545}$, which is in good agreement with previously-reported results provided that the $\sim Pr^{-0.82}$ dependence \citep{Li2021} and $\Gamma$ dependence \citep{VanDerPoel2012} of $Re$ are taken into account.

Our HC scalings are in relatively good agreement with the theoretical predictions $Nu^{\chi}_{HC}\sim Ra_L^{0.2}$ and $Re\sim Ra_L^{0.4}$ based on laminar conditions, with $Re$ usually based on the peak velocity within the HC boundary layer \citep{Rossby1965,Hughes2008,Sheard2011} instead of the mean kinetic energy density. The slightly higher exponent for the Nusselt number $a_{HC}=0.224>0.2$ is likely due to the unsteadiness of the horizontal flow given our relatively large $Ra_L$ \citep{Sheard2011}, while remaining substantially smaller than the upper bound 1/3, which is the exponent of the \textit{ultimate regime} \citep{Rocha2020a}. Again, we note that the small dependence of $Nu^{\chi}_{HC}$ with $\Gamma$, which has not been observed previously \citep{Sheard2011}, comes from the normalization, which is based on the diffusive solution rather than on a purely dimensional argument (see equation \eqref{eq:nuhc}).

\section{Dissipation of temperature variance and diffusive solution}\label{appD}

Recently, \citet{Rocha2020} showed that a compelling definition of the Nusselt number of horizontal convection should involve the dissipation of temperature (or buoyancy $b=-T$) variance, $\chi = \overline{\langle |\NA T|^2 \rangle}$, because it is related to the vertically-averaged horizontal heat flux set up by the imposed heterogeneous temperature profile at (here) $z=1$. They further showed that $\chi =\overline{\langle T(z=1)\p_z T(z=1) \rangle_x}$, such that it can be evaluated through a line integral, which is more tractable in laboratory experiments and numerical simulations than a volume integral as required by $\overline{\langle |\NA T|^2 \rangle}$. Here we derive a similar result for the case with geothermal heating. Multiplying \eqref{eq:a13} by $T$ and rearranging yields
\ba{}\label{eq:budg1} 
\p_t \lp \f{T^2}{2} \rp + |\NA T|^2 = - \NA\cdot\lp \frac{\u T^2}{2} \rp + \NA\cdot\lp T\NA T \rp. 
\ea
The first term becomes zero when time averaging at statistical steady state, while the third term becomes zero when performing a volume average because of the no-slip condition on the walls. Thus, 
\ba{}\label{eq:budg2}
\chi=\overline{\langle|\NA T|^2\rangle} = \overline{\langle\NA\cdot\lp T\NA T \rp\rangle} = \overline{\langle T(z=1)\p_z T(z=1) - T(z=0)\p_z T(z=0) \rangle_x}  > 0,
\ea
where the last equality is obtained after integrating in $z$ and enforcing no-flux conditions on the side walls. Our expression for $\chi$ differs from that of \citet{Rocha2020} because of the second term on the right-hand-side of equation \eqref{eq:budg2}, which is non-zero here because our derivation takes into account the heat flux ($\p_z T=-1$) on the bottom boundary.

The diffusive solution in HC, which we derive in terms of dimensionless variables in this section, can be readily obtained using the method of separation of variables. For the surface temperature profile of equation \eqref{eq:bcs2}, the diffusive solution without geothermal heating ($\p_z T=0$ at $z=0$) is
\ba{} 
T_{\mathrm{diff}} = \f{\Lambda\Gamma \sin \lp \f{\pi x}{\Gamma} \rp \cosh \lp \f{\pi z}{\Gamma} \rp }{2 \cosh \lp \f{\pi}{\Gamma} \rp} ,
\ea
while, with geothermal heating ($\p_z T=-1$ at $z=0$), we simply add a linearly-varying vertical profile, i.e.
\ba{}\label{eq:tdiff}
T_{\mathrm{diff}} = \f{\Lambda\Gamma \sin \lp \f{\pi x}{\Gamma} \rp \cosh \lp \f{\pi z}{\Gamma} \rp }{2 \cosh \lp \f{\pi}{\Gamma} \rp} + 1-z.
\ea
{The denominator in equation \eqref{eq:nuhc} based on equation \eqref{eq:tdiff} then reads
\ba{}\label{eq:budg3} 
\chi_{\mathrm{diff}} = \langle \p_z T_{\mathrm{diff}}(z=1) T_{\mathrm{diff}}(z=1) - \p_z T_{\mathrm{diff}}(z=0) T_{\mathrm{diff}}(z=0)\rangle_x = \f{\pi\Lambda^2\Gamma}{8}\tanh\lp \f{\pi}{\Gamma}\rp +1,
\ea
which tends asymptotically to $\pi^2\Lambda^2/8+1$ for large $\Gamma$ with the $+1$ term being inherited from geothermal heating (which is thus discarded for the calculation of $Nu^{\chi}_{HC}$ in pure HC simulations). Importantly, geothermal heating has a negligible effect on $\chi_{\mathrm{diff}}$ only if $\Lambda\gg 1$.}

\section{Nusselt numbers of horizontal convection}\label{appE}

\begin{figure}
\centering
\includegraphics[width=0.88\textwidth]{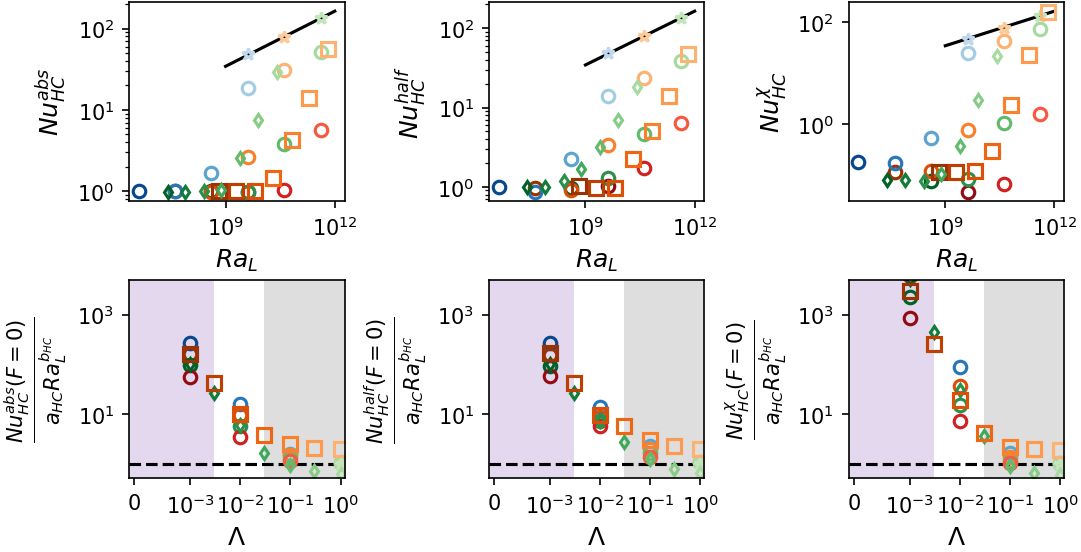}
\includegraphics[width=0.11\textwidth]{grille-de-lecture-des-symboles.png}
\vspace{-0.1in}\caption{{The top row shows the three Nusselt numbers defined in equations \eqref{eq:nuabshc}-\eqref{eq:nuhc} as functions of $Ra_L$ with the solid black line showing the scaling law derived from the $Nu^{\chi}_{HC}$ data obtained for pure HC simulations. The bottom row shows the same Nusselt numbers divided by the scaling law and renormalized by the diffusive solution derived without geothermal heating ($F=0$); e.g. $Nu^{\chi}_{HC}(F=0)=Nu^{\chi}_{HC}\times\chi_{\mathrm{diff}}/\chi_{\mathrm{diff}}(F=0)$.}}
\label{figappE}
\end{figure}

{Figures \ref{figappE}(a)-(c) show $Nu^{\text{abs}}_{HC}$, $Nu^{\text{half}}_{HC}$ and $Nu^{\chi}_{HC}$ (defined in equations \eqref{eq:nuabshc}-\eqref{eq:nuhc}) as functions of $Ra_L$. Pure HC simulation results are indistinguishable from the scaling law (solid line) derived from the $Nu^{\chi}_{HC}$ data, which means that the heat transfer efficiency by horizontal convection is the same for all three definitions of the Nusselt number in the HC-dominating limit. Differences exist between RBH simulation results (compare e.g. light blue, orange and green circles in figures \ref{figappE}(b) and \ref{figappE}(c)). However, for $\Lambda\gg 10^{-2}$, these differences are due to the diffusive normalization (i.e. appearing at the denominator), whose dependence on geothermal heating remains significant for $\Lambda\leq O(1)$ (unlike the numerator) and depends on the choice of Nusselt number definition. When renormalizing the numerator in the definition of each Nusselt number by the diffusive solution \textit{without} geothermal heating (i.e. considering $F=0$), those differences (at large $\Lambda$) are removed, as can be seen from figures \ref{figappE}(d)-(f) in which all renormalized Nusselt numbers divided by the scaling law $a_{HC}Ra_L^{b_{HC}}$ converge toward unity. Note that differences due to aspect ratio $\Gamma$ are also due to the diffusive normalization, as explained at the end of \S~\ref{sec:nuss}.}

\bibliographystyle{jfm}
\bibliography{export}

\end{document}